\documentclass[iop,floatfix]{emulateapj}
\usepackage{apjfonts}

\shorttitle{SN\,2010U}
\shortauthors{Czekala et al.}

\usepackage{hyperref}

\newcommand{\obj}{SN\,2010U}
\newcommand{\pc}{P~Cygni}
\def\ra#1#2#3{#1$^{\rm h}$#2$^{\rm m}$#3$^{\rm s}$}
\def\dec#1#2#3{$#1^\circ#2'#3''$}
\newcommand{\kms}{\;\textrm{ km }\textrm{s}^{-1}}
\newcommand{\HET}{February 7.30}
\newcommand{\NOT}{February 12.21}
\newcommand{\GMOS}{February 21.51}
\newcommand{\maxdate}{6.27}

\begin{document}

\title{The Unusually Luminous Extragalactic Nova SN\,2010U}

\author{
Ian~Czekala\altaffilmark{1},
E.~Berger\altaffilmark{1},
R.~Chornock\altaffilmark{1},
A.~Pastorello\altaffilmark{2},
G.~H.~Marion\altaffilmark{1,3}, 
R.~Margutti\altaffilmark{1},
M.~T.~Botticella\altaffilmark{4},
P.~Challis\altaffilmark{1},
M.~Ergon\altaffilmark{5}, 
S.~Smartt\altaffilmark{2}, 
J.~Sollerman\altaffilmark{5},
J.~Vink\'{o}\altaffilmark{6},
J.~C.~Wheeler\altaffilmark{3}
}

\email{iczekala@cfa.harvard.edu}

\altaffiltext{1}{Harvard-Smithsonian Center for Astrophysics, 60
Garden Street MS 10, Cambridge, MA 02138}
\altaffiltext{2}{Astrophysics Research Centre School of Mathematics
and Physics, Queen's University Belfast, Belfast BT7 1NN, Northern
Ireland UK}
\altaffiltext{3}{The University of Texas at Austin Department of
Astronomy, RLM 5.208, Austin, TX 78712-1081}
\altaffiltext{4}{INAF-Osservatorio astronomico di Capodimonte Salita
Moiariello, 16 80131 Napoli, Italy}
\altaffiltext{5}{The Oskar Klein Centre, Department of Astronomy,
AlbaNova, Stockholm University, 10691 Stockholm, Sweden}
\altaffiltext{6}{Department of Optics and Quantumelectronics
University of Szeged, Szeged, Hungary}

\begin{abstract}
We present observations of the unusual optical transient \obj, including spectra taken 1.03 days to 15.3 days after maximum light that identify it as a fast and luminous \ion{Fe}{2} type nova. Our multi-band light curve traces the fast decline ($t_2 = 3.5 \pm 0.3$ days) from maximum light ($M_V = -10.2 \pm 0.1$ mag), placing \obj\ in the top 0.5\% of the most luminous novae ever observed. We find typical ejecta velocities of $\approx 1100 \kms$ and that \obj\ shares many spectral and photometric characteristics with two other fast and luminous \ion{Fe}{2} type novae, including Nova LMC 1991 and M31N-2007-11d. For the extreme luminosity of this nova, the maximum magnitude vs. rate of decline relationship indicates a massive white dwarf progenitor with a low pre-outburst accretion rate. However, this prediction is in conflict with emerging theories of nova populations, which predict that luminous novae from massive white dwarfs should preferentially exhibit an alternate spectral type (He/N) near maximum light.  
\end{abstract}


\section{Introduction}
Unprecedented areal and temporal coverage of the sky from dedicated surveys and amateur observers has greatly amplified the discovery rate of unusual optical transients.  Surveys such as Pan-STARRS, the Palomar Transient Factory, and the Catalina Real-Time Transient Survey have demonstrated the wealth of data that will be common in the era of the Large Synoptic Survey Telescope (LSST).  In particular, a previously sparse regime of transient phase-space between classical novae ($M_{V{\rm,\,peak}}\sim -8$ mag; \citealt{be08}) and supernovae ($M_{V{\rm,\,peak}}\sim -18$ mag; \citealt{fil97}) is now being populated with an increasing number of transients.  These objects are quite diverse in their properties and may shed light on a wide range of explosion and eruption physics.

In recent years, objects like SN\,2008S and NGC\,300\,OT \citep{pkt+08,bps+09,bsc+09, tps+09, spk+12} and other luminous blue variables (LBVs) \citep{hd94,pbt+10,sls+11} have been subjected to intense scrutiny. These intermediate luminosity optical transients (ILOTs; also referred to as SN impostors and luminous red novae) might be the eruption of a dust-enshrouded massive star and promise to lend great insight into the late stages of massive stellar evolution or other poorly understood stellar physics. Because the phase space these eruptions inhabit is crowded with fundamentally different transient systems, it is important for future transient discovery scrutinize this region with intensive spectroscopic and photometric followup to distinguish these explosions from other more traditional explosions, such as classical novae or supernovae, which to fall in this region of phase space would qualify them as remarkable in their own right.

Classical novae (CNe) are binary systems where there is mass transfer from a (possibly evolved) secondary through the L1 Lagrange point to a degenerate white dwarf (WD). When enough material has accreted to obtain critical temperature and density, nuclear burning begins. The $p-p$ chain gives way to CNO reactions, which drive convection. The amount of energy deposited by the $\beta^+$ unstable nuclei then drives a radiative wind. Because degenerate matter on the surface of the WD has a equation of state independent of temperature, these reactions proceed in a runaway fashion until the Fermi temperature is reached and the surface layers of the white dwarf begin to function as an ideal gas sensitive to temperature and finally expand.  This expansion speed can easily reach escape velocity and the radiation pressure ejects a shell of material \citep{war03}. 

The inferred classical nova rate in the Milky Way is $\sim 35\;{\rm yr}^{-1}$ \citep{dbk+06}, however interstellar extinction and selection effects limit the number of observed novae. The mean absolute magnitude of novae is $M_V \approx -7.5$ mag, and of nearly a thousand novae on record, less than 10 reached peak absolute magnitude brighter than $M_V = -10.0$ mag \citep{srq+09}.  

Here we present the detailed photometric and spectroscopic observations of \obj. We show that \obj\ is clearly super-Eddington at maximum light and identify it as a close spectroscopic analog to other super-Eddington novae. We compare \obj\ to the general nova population and recognize it as one of the most luminous and fast declining novae discovered to date. These characteristics of \obj\ make it a valuable object to study in the context of outburst models and progenitor studies of luminous novae.

\section{Discovery and Reduction}
\begin{figure}
\begin{center}
\includegraphics[width=0.47\textwidth]{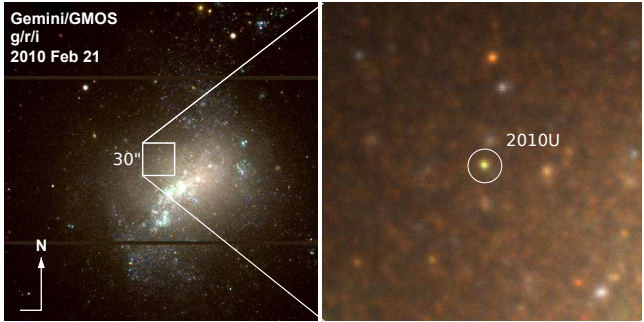} 
\caption{The field of \obj\ in NGC\,4214, observed on 2010 February
21.54 with GMOS on Gemini North.  The composite 3-color image combines
observations with $gri$ filters. The inset on the right shows \obj\
about two weeks after maximum light.}
\label{fig:galaxy}
\end{center}
\end{figure}

\citet{nk10} discovered \obj\ in NGC\,4214 on 2010 February 5.63 UT (UT dates are used throughout this paper) and was subsequently observed by several amateur astronomers that night.  \obj\ is located at RA=\ra{12}{15}{41.06}, Dec=\dec{+36}{20}{02.9} (J2000), about $20"$ east and $27"$ north of the center of NGC\,4214 \citep{nk10} (Figure~\ref{fig:galaxy}).  We use the distance modulus of $m - M = 27.41\pm 0.03$ mag \citep{dws+09} for NGC\,4214 and correct all magnitudes for Galactic reddening of $E(B-V) = 0.02$ mag using the dust maps of \citet{sfd98}.  Observations by Itagaki provide a pre-explosion limit of 18.8 mag (unfiltered) on 2010 January 24.74.  \citet{hpr+10} determined that \obj\ was initially mis-classified as a supernova, and is in fact a luminous and fast classical nova.  They conclude that \obj\ reached a peak absolute magnitude of $M_R = -10.5$ mag and faded two magnitudes on a timescale of $t_2\approx 15$ d. They use a distance modulus of $m - M = 27.53$, while our distance modulus determination is more recent. Adopting our distance modulus, the peak absolute magnitude using the results of \citet{hpr+10} is $M_R = -10.4$ mag.

\subsection{Photometry}
\label{sec:phot}
We initiated a multi-band photometric follow-up campaign of \obj\
starting on 2010 February 6.98 using the 2-m Liverpool Telescope (LT;
\citealt{ssr+04}) with RatCam; the 8-m Gemini North Telescope (GN)
with Gemini Multi-Object Spectrograph (GMOS; \citealt{had+04}); and
the 2.56-m Nordic Optical Telescope (NOT; \citealt{da10}) with the
Andalucia Faint Object Spectrograph and Camera (ALFOSC).  We also
collected photometry from amateur astronomers K.~Itagaki, T.~Yusa,
J.~Brimacombe, and J.~Nicolas, who kindly provided us their unfiltered
discovery images from 2010 February 5.65 to 13.03, which captured the
rise and peak of \obj.

We bias-subtracted and flat-fielded all images using standard techniques in IRAF, and determined instrumental magnitudes using PSF-fitting of the source.  We obtained nightly zero-points by observing a number of standard fields from the \citet{lan92} catalog. We improved the calibration of individual magnitudes of the transient through comparison with the average magnitudes of a local stellar sequence in the field of \obj\ established during selected photometric nights.

Observations from the LT used Landolt $B$- and $V$-band and Sloan $r^\prime$- and $i^\prime$-band, but were calibrated to Landolt standards in the Vega system.  To place all fluxes on the same system, we transformed these measurements to the AB system using offsets derived from ${\tt pysynphot}$\footnote{\url{http://stsdas.stsci.edu/pysynphot/}} of $-0.115$, $0.000$, $+0.142$ and $+0.356$ mag, respectively.  Unfiltered observations with the NOT, as well as amateur observations, were unfiltered but initially calibrated to Vega $R$-band, so we transformed these measurements to AB using an offset of $+0.183$ mag.  Since GN observations were initially calibrated to the AB system no transformation was necessary, and all magnitudes quoted in this paper are AB unless otherwise noted.  The quantum efficiency curves of the instrumental configurations used by the amateur astronomers peak around $6000-6200$ \AA, so the unfiltered magnitudes were scaled to match the $r^\prime$-band photometry.  In the case of the observation by J.~Brimacombe, however, the transmission curve of his luminance filter\footnote{Transmission function at \url{http://www.sbig.com/sbwhtmls/}\\\url{announcement_baader_narrowband_f2.htm} } peaks at 5500 \AA\ and sharply declines outside the range $4200-6800$ \AA, making calibration to $V$-band most appropriate. We summarize the optical photometric measurements in Table~\ref{table:phot} and present the light curve in Figure~\ref{fig:lightcurve}.
 
We also observed \obj\ with the \emph{Swift} satellite \citep{gcg+04} on 2010 March 3.82 with the X-Ray Telescope (XRT; \citealt{bhn+05}) and the UV/Optical Telescope (UVOT; \citealt{rkm+05}).  We did not detect any X-ray or UV/optical emission coincident with the location of the source.  A previous \emph{Swift}/XRT observation of NGC\,4214 on 2007 March 26.50, which included the field of \obj, showed no activity coincident with the source location. We analyzed all \emph{Swift} data with the ${\tt Heasoft-6.11}$ software package and corresponding calibration files, applying standard screening and filtering criteria. We reduced XRT data with the ${\tt xrtpipeline}$ and determined $3\sigma$ upper limits with the ${\tt sosta}$ task in the ${\tt ximage}$ suite using a $5"$ radius aperture; see Table~\ref{table:swift_xrt}.  We processed UVOT with the standard UVOT data reduction pipeline \citep{pbp+08} and determined $3\sigma$ upper limits with a $5"$ radius aperture; see Table~\ref{table:swift_uvot}.

\subsection{Spectroscopy}
We obtained three low resolution optical spectra of \obj\ using the Marcario Low-Resolution Spectrograph (LRS, \citealt{hnm+98}) on the Hobby-Eberly Telescope (HET; \citealt{rab+98}), ALFOSC on NOT, and GMOS on GN. We reduced the NOT spectrum using the QUBA pipeline \citep{vfb+11}, implemented in IRAF, and the HET and GMOS spectra using standard tasks in IRAF. We observed all targets at low airmass ($\lesssim 1.2$) with the slit was aligned to the parallactic angle, and flux-calibrated each spectrum using a spectrophotometric standard star observed at a similar airmass.  All spectra were wavelength-calibrated by comparison with Helium-Neon-Argon arc lamps.  We summarize spectroscopic measurements and instrumental configurations in Table~\ref{table:spectroscopy}.  We analyzed the resulting 1-d spectroscopic data in IRAF using ${\tt onedspec}$ tasks and the Scipy Python packages \citep{jop+01}.

\section{Results}
Complete photometric and spectroscopic coverage of \obj\ confirms the findings of \citet{hpr+10}: \obj\ is a luminous classical nova, exhibiting a rapid optical decline and evolution from an optically thick spectrum dominated by hydrogen and iron emission lines to an optically thin nova spectrum entering the nebular stage.  \obj\ is not a supernova nor the eruption of a massive star because of its modest ejecta velocities ($\approx 1100\kms$) and rapid optical decline and spectral evolution. Supernovae typically exhibit much higher expansion velocities ($\gtrsim 10^4\kms$) \citep{fil97} and while LBV eruptions exhibit a range of expansion velocities ($\sim 200 - 2000\kms$; \citealt{sls+11}) and strong hydrogen Balmer emission, the presence of CNO element lines and rapid optical and spectral evolution of \obj\ strongly indicate that it was a classical nova and not an LBV.

\subsection{Optical Photometric Evolution}

\begin{figure*}[htb]
\begin{center}
\includegraphics[width=0.9\textwidth]{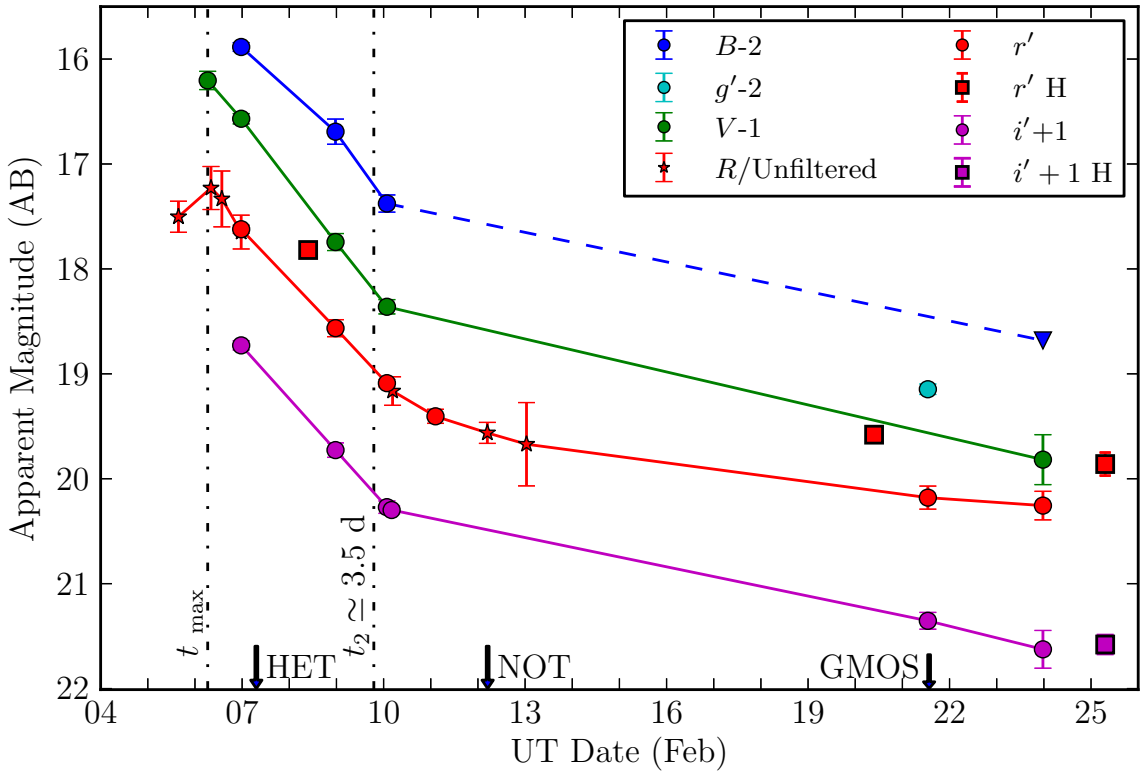}
\caption{Light curves of \obj\ corrected for Galactic extinction \citep{sfd98} and plotted in AB magnitudes.  $B$-, $V$-, and $i^\prime$-band data points are offset for clarity, and unfiltered amateur observations are calibrated to $R$-band.  We adopt the date of $V$-band maximum, 2010 February \maxdate, as the date of maximum light for \obj\ at apparent magnitude of $17.28$ mag.  The $g^\prime$-band measurement is distinct from nearby $B$- and $V$-band measurements because the differences in the passbands yield significant deviations due to the presence of emission lines (Figure~\ref{fig:overall_spectra}).  Photometry from \citet{hpr+10} is converted to the AB system, corrected for extinction, and included in this light curve.  We find a $\sim 0.3$ mag difference in $r^\prime$-band, which is the source of disagreement in our determination of $t_2$. However, we note that the very rapid decline is also apparent in $B$-, $V$-, and $i^\prime$-band.}
\label{fig:lightcurve}
\end{center}
\end{figure*}

\obj\ evolved rapidly after its discovery on 2010 February 5.63.  Our light curve is well sampled near maximum light in $r^\prime$-band, and the transient is seen to rise $\approx 0.25$ magnitude from discovery to maximum.  Although the rise of \obj\ is not captured in $V$-band, the first measurement at $17.28$ mag is contemporaneous with the measurement of maximum light in $r^\prime$-band.  For the purpose of comparison with previous events, we adopt the date of $V$-band maximum, 2010 February \maxdate, as the date of maximum light.  After maximum, \obj\ rapidly declined with a linear slope in magnitude space, at first steeply and then becoming more gradual after 2010 February 10 (Figure~\ref{fig:lightcurve}).  We followed the light curve of the transient until 2010 February 24. \citet{hpr+10} followed the transient until 2010 March 18.40, reporting a continued steady decline.

The photometric evolution of classical novae is typically parameterized by the time to decline by two magnitudes from maximum light, $t_2$.  Several studies have shown that $B$- or $V$-band are most appropriate to measure $t_2$ because H$\alpha$ emission complicates measurements in $r^\prime$ band \citep{sdh+11,be08}.  We measure $t_2$ by adopting the $V$-band maximum (2010 February \maxdate), and then linearly interpolating between two $V$-band measurements at 2010 February 8.98 and 2010 February 10.07 in magnitude space, which gives a result that is accurate to $\pm 0.13$ days.  The uncertainty in the date of maximum light derives from the assumption that maximum light in $V$-band corresponds with maximum light in $r^\prime$ band, and therefore we have captured the peak of the light curve to $\pm 0.3$ d.  We find that \obj\ underwent a fast decline, with a $V$-band maximum of $M_V=-10.2\pm 0.1$ mag and $t_2=3.5\pm 0.3$ days.  Our determination of absolute magnitude and $t_2$ are in contrast to the results of \citet{hpr+10}, who derive $M_{R,{\rm max}}\approx -10.5$ mag and $t_2\approx 15$ d.  This is primarily due to the smaller distance modulus adopted here, and our better sampling of the light curve in the range $t_{\rm max}$ to $t_{\rm max}+15$ d.  In addition, a fast decline is evident in $B$-, $V$-, and $i^\prime$-band as well.  The rise time of \obj\ from quiescence to maximum light remains unconstrained due to the comparatively shallow upper limit ($m_{r^\prime}\approx 18.8$ mag) on January 24.74 UT and a large gap before discovery.

\begin{figure}
\begin{center}
\includegraphics[width=0.48\textwidth]{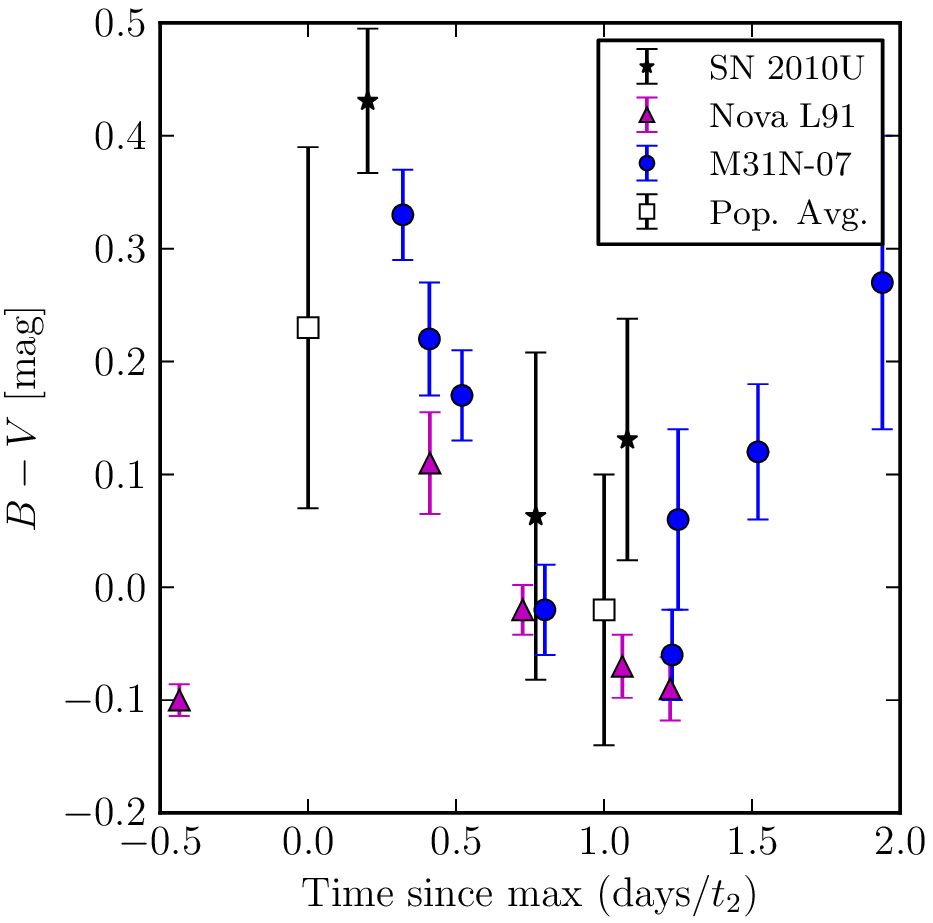}
\caption{Colors of \obj, L91, M31N, and the general nova population \citep{vy87} in units of $t_2$ after correction for Galactic reddening.  A comparison of \obj\ to L91, another luminous and fast nova, assuming that any color difference is due to intrinsic host galaxy extinction, indicates that \obj\ might suffer as much as $E(B-V)\approx 0.2$ mag of additional intrinsic extinction, raising the peak luminosity to $M_V\approx -10.9$ mag.  For the purpose of comparison, all colors are given in Vega magnitudes.}
\label{fig:colors}
\end{center}
\end{figure}

We compare the colors of \obj\ to those of other fast and luminous novae and the general nova population to determine if there is intrinsic host galaxy extinction.  In Figure~\ref{fig:colors}, the colors of \obj\ are plotted against another fast and luminous nova, Nova LMC 1991 (hereafter L91), and the average colors of the nova population. \citet{vy87} find that of 7 novae at maximum light, $(B - V)_{\rm avg}^{\rm max} = 0.23\pm 0.06$ mag, with a dispersion $\sigma_{B-V}\lesssim 0.16$ mag.  They also find that at $t_2$, 13 novae are found to have an intrinsic color $(B - V)_{\rm avg}^{t_2} = -0.02 \pm 0.04$ mag, with a dispersion $\sigma_{B-V}\lesssim 0.12$ mag.

It is interesting to compare \obj\ to L91 and speculate that any color difference might be due to intrinsic host galaxy extinction.  For \obj\, near maximum light $(B - V) = 0.43 \pm 0.06$ mag and near $t_2$ $(B - V) = 0.13 \pm 0.11$ mag.  If we were to assume that \obj\ has the same intrinsic colors as L91, then \obj\ might suffer as much as $E(B-V) \approx 0.2$ mag of additional intrinsic extinction, which would raise its peak brightness to $M_V \approx -10.9$ mag, making it the most luminous classical nova on record.  However, \obj\ could also simply be intrinsically redder than L91.

\citet{hpr+10} assume the $V - R \approx 1.1$ mag color of V1500 Cyg, another luminous nova, to infer $M_V \approx - 9.4$ mag for \obj.  However, we measure $V - r^\prime = 0.1$ mag (Vega), suggesting that \obj\ was not as red as V1500 Cyg.  We conservatively assume no intrinsic host galaxy extinction for all further analysis.

\subsection{X-ray}

The \emph{Swift} XRT and UVOT observed the location of \obj\ 25.6 days after maximum optical light and did not detect the source (Tables~\ref{table:swift_xrt}~\&~\ref{table:swift_uvot}), placing a 3$\sigma$ upper limit of $9.1 \times 10^{-3}~{\rm ct s}^{-1}$ for the X-rays in the 0.3-10.0~keV energy band. Using the relationship derived in \citet{go09} to convert optical extinction $A_V$ into hydrogen column density $N_H$, we obtain $N_H ({\rm cm}^{-2}) = (2.21 \pm 0.09) \times 10^{21} A_V\;({\rm mag}) = 1.60 \times 10^{20}\;{\rm cm}^{-2}$. Using the \emph{Chandra} X-ray Center's \emph{Portable, Interactive Multi-Mission Simulator} (PIMMS)\footnote{\url{http://cxc.harvard.edu/toolkit/pimms.jsp}}, and assuming a spectrum for the nova X-ray emission, we convert count rates into flux limits by assuming a spectrum for the nova.

\citet{sno+11} present a compilation of 52 Galactic and Magellanic Cloud CNe and recurrent novae (RNe) observed with the \emph{Swift} XRT. X-ray studies of CNe have identified two different emission components, a hard X-ray component and a soft X-ray component. The fastest optically declining novae (as measured by $t_2$) usually have an early hard X-ray phase, while the slower novae do not.  The hard X-ray emission may originate from shocks between the fast moving ejecta and pre-existing circumstellar material, and typically is hard thermal bremsstrahlung ($T \sim 6 \times 10^7 - 1 \times 10^8)$~K, low luminosity $(\sim 10^{34}\;{\rm erg\;s}^{-1}$ \citep{bko98}, and of shorter duration than the soft X-ray phase \citep{sno+11}.  The soft phase begins when the nova shell becomes optically thin and the photosphere of the nova recedes to the surface of the hot WD, with blackbody emission at $T = 2-8 \times 10^5$~K \citep{sno+11}. This emission lasts as long as nuclear reactions continue on the surface of the white dwarf. \citet{sno+11} find that the Super Soft X-ray phase begins and ends sooner for fast novae (as measured by $t_2$) than for slow novae and that novae with slower expansion velocities will enter the Super Soft state later but emit X-rays for longer. The correlation between Super Soft X-ray turn off time and $t_2$ has significant scatter \citep{hk10,sno+11}, but if for \obj\ $t_2 = 3.5 \pm 0.3$ d, then the turn-off time would be lower than 60 d and possibly as low as 10 d. 

Adopting a temperature of $k T = 5$ keV for the hard component of \obj\ would place a $3\sigma$ upper limit on the X-ray luminosity of of $L_X = 1.6 \times 10^{39}\;{\rm erg\;s}^{-1}$, while adopting a temperature of $k T = 60$~eV for the soft component of \obj\ would place a limit of $L_X = 2.6 \times 10^{38}\;{\rm erg\;s}^{-1}$. While neither of these limits are strong constraints, the upper limit on the Super Soft emission approaches the X-ray luminosities of some novae on record. \citet{sno+11} found that for Nova\,V407\,Cyg the blackbody luminosity of the Super Soft emission was $L_X = 9.3 \times 10^{37}\;{\rm erg\;s}^{-1}$ at 27 d after optical maximum, and that nuclear burning on the surface of the WD occurred from eruption until about 30 d after optical maximum, meaning that \obj\ could not have been much brighter in X-rays than Nova\,V407\,Cyg. The deeper pre-explosion observation on 2007 March~26.50 of the field of \obj\ placed a 3$\sigma$ upper limit of $6.7 \times 10^{-3}~{\rm ct s}^{-1}$, providing a weak upper limit on the luminosity of the nova system in quiescence.  Adopting a temperature of $k T = 60$ eV for a soft quiescent spectrum of \obj\ would place a limit of $L_X = 2.1 \times 10^{38}\;{\rm erg\;s}^{-1}$.

\subsection{Bolometric Flux Evolution}

\begin{figure}
\begin{center}
\includegraphics[width=0.48\textwidth]{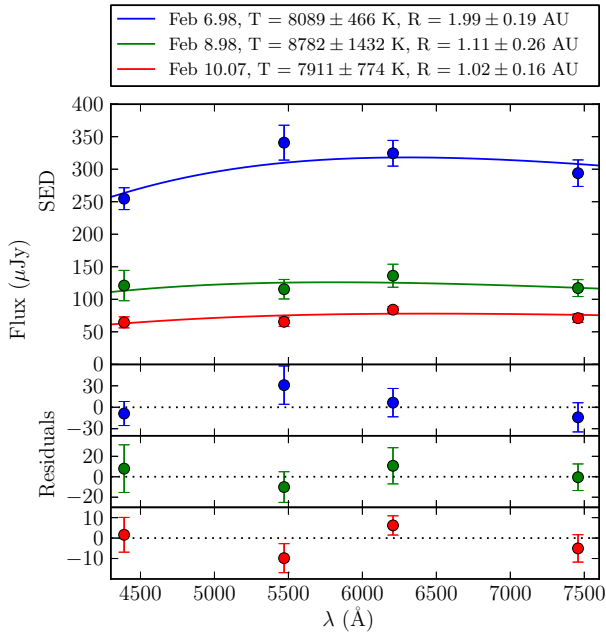}
\caption{Best-fit radius and temperature parameters for each epoch. Within the errors, the temperature of the photosphere remains constant over the three epochs while the radius of the photosphere recedes to $\approx 1.0$~AU as the envelope becomes optically thin.} 
\label{fig:sed}
\end{center}
\end{figure}

During the early evolution of the light curve near maximum light $(t \lesssim 4$ d), the ejected shell of \obj\ is still optically thick and we can fit the spectral energy distribution (Figure~\ref{fig:sed}) with a spherical blackbody function.  We use the photometry from 2010 February 6.98, 8.98, and 10.07.  Effective wavelengths for these filters were determined using ${\tt pysynphot}$ and the HET and NOT spectra, yielding $\lambda_{B, {\rm eff}} = 4387$ \AA, $\lambda_{V, {\rm eff}} = 5468$ \AA, $\lambda_{r^\prime, {\rm eff}} = 6202$ \AA, and $\lambda_{i^\prime, {\rm eff}} = 7463$ \AA.

We use $\chi^2$ minimization to find the best-fit parameters of radius and temperature, shown in Figure~\ref{fig:sed}. For the spectral energy distribution (SED) nearest maximum light (2010 February 6.98), we obtain a photospheric temperature of $T=8090 \pm 470$~K and a radius of $R=1.99\pm0.19$~AU. Within the errors, the temperature of the photosphere remains constant for the following two epochs, while its radius recedes to $\approx 1.0$~AU, indicating that the envelope becomes optically thin.  This temperature fits well with the typical $T_{\rm eff} \leq 10^4$~K derived for novae at visual maximum \citep{wil92}.

If we combine the expansion velocity measured from spectral lines (see \S~\ref{sec:spectral_evolution}) with the radius of the photosphere and assume ballistic expansion, we can estimate the time since explosion.  We determine that time from explosion to 2010 February 6.98 was $t = 3.06\pm 0.40$ d.  This suggests a rapid rise to maximum, but is otherwise consistent with the observations since it is uncertain how accurately the spectral line widths probe the bulk ejecta velocity, because the lines may be formed in a wind.

\begin{figure}
\begin{center}
\includegraphics[width=0.48\textwidth]{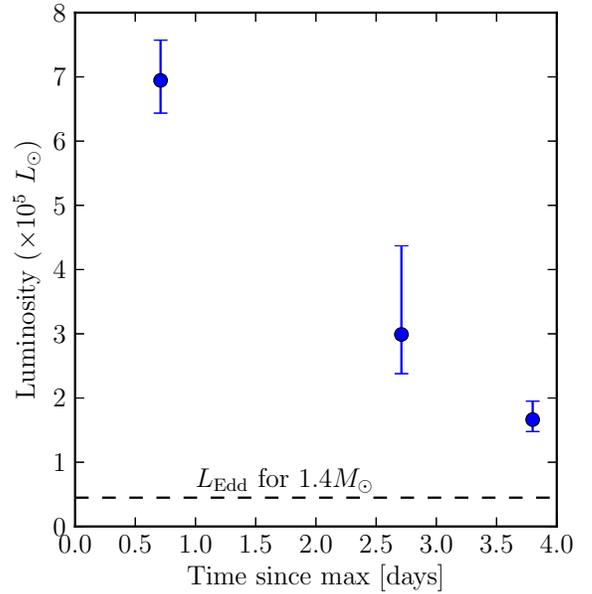}
\caption{Bolometric luminosity of \obj\ determined from fits of a spherical blackbody to photometry (See Figure~\ref{fig:sed}). Shortly after maximum light (+0.71~d), we find that \obj\ is clearly super-Eddington with luminosity $L=(2.71 \pm 0.22) \times 10^{39}\;{\rm erg\;s}^{-1}$. Over the next three days the luminosity declines by a factor of 4.}
\label{fig:super_eddington}
\end{center}
\end{figure}

Using the best fit parameters, we estimate the blackbody luminosity of the photosphere. These luminosities are plotted in Figure~\ref{fig:super_eddington} along with the Eddington Luminosity for a $1.4~M_\odot$ white dwarf, $L_{\rm Edd} = 1.75\times 10^{38}\;{\rm erg\;s}^{-1}$, calculated using a 100\% ionized atmosphere and Thompson scattering opacity. On 2010 February 6.98, we find that $L=(2.71 \pm 0.22) \times 10^{39}\;{\rm erg\;s}^{-1}$, and the luminosity declines by a factor of 4 over the next three days. \obj\ is clearly super-Eddington for at least the 4 days near maximum light, in agreement with determination by \citet{hpr+10}.

The super-Eddington luminosity of \obj\ is similar to that of L91, where model-atmosphere fitting to UV and optical data by \citet{sss+01} determined that it remained super-Eddington from $-5< t<8$ d around maximum light.  They find a peak bolometric flux of $L = (2.6 \pm 0.3) \times 10^{39}\;{\rm erg\;s}^{-1}$ with $R = 0.7$~AU and $T = 1.3 \times 10^4$~K at maximum light.  \citet{sss+01} determine from their model that the radiative forces are ten times the gravity forces for the entire atmosphere, thus the ``atmosphere'' should appear as a radiatively driven wind.  \citet{sha01} suggests that a clumpy but porous photosphere would enable steady-state super-Eddington luminosities to persist for an extended period.

\citet{sss+01} determine from their model that the radiative forces are ten times the gravity forces for the entire atmosphere, thus the ``atmosphere'' should appear as a radiatively driven wind. \citet{sha01} suggests that a clumpy but porous photosphere would enable steady-state super-Eddington luminosities to persist for an extended period. 

\subsection{Spectroscopic Evolution}
\label{sec:spectral_evolution}

\begin{figure*}[htb]
\begin{center}
  \includegraphics[width=0.82\textwidth]{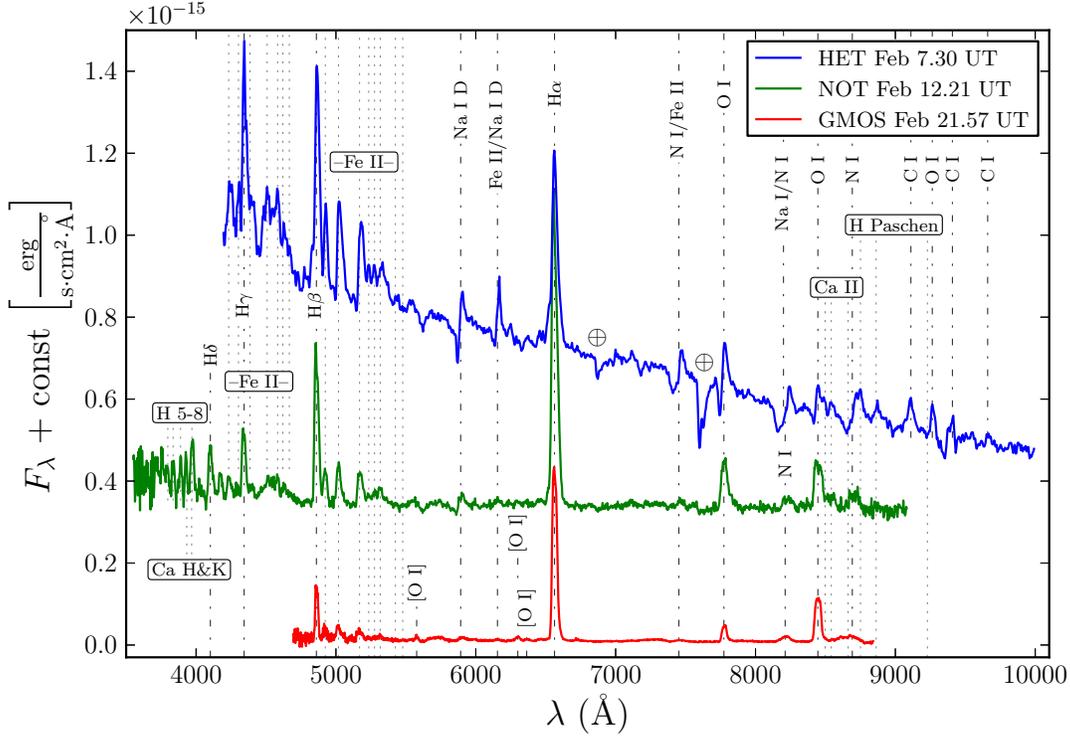}
\caption{Multi-epoch spectra of \obj\ capturing the post-maximum ``iron-curtain'' stage (HET), ``pre-nebular'' stage (NOT), and the onset of the nebular stage (GMOS). Spectra were de-redshifted to match the [\ion{S}{2}] $\lambda6716.44$, $\lambda6730.82$ \AA\ emission lines from the host galaxy ($z = 0.00087$). Strong P Cygni profiles are present at early times (see Figure~\ref{fig:pcygnis}) but quickly fade into the continuum.}
\label{fig:overall_spectra}
\end{center}
\end{figure*}

\citet{hpr+10} published a spectrum of \obj\ 14 days after maximum light, noting the presence of H$\alpha$, H$\beta$, and \ion{O}{1}$\lambda\lambda7774$.  They emphasize that this spectrum does not resemble that of a supernova nor any intermediate luminosity optical transients such as SN\,2008S nor NGC\,300\,OT.

Our three epochs of spectroscopy trace the classical nova spectral evolution of \obj\ from 1.03 d after maximum light (2010 \HET) to 15.30 d after maximum light (2010 \GMOS) (Figure~\ref{fig:overall_spectra}). The earliest spectrum exhibits strong emission lines of (in decreasing strength) the hydrogen Balmer series, \ion{Fe}{2}, \ion{Na}{1}, \ion{O}{1}, \ion{N}{1}, and \ion{C}{1}.  Spectra were de-redshifted to match the [\ion{S}{2}] $\lambda6716.44$, $\lambda6730.82$ emission lines from the host galaxy ($z = 0.00087$).  The radial velocity of \obj\ is $-260\kms$, while the NED redshift of NGC\,4214 is $-290\kms$. This $\Delta v \approx -30 \kms$ is consistent with the internal motions of the galaxy.

\begin{figure}
\begin{center}
\includegraphics[width=0.45\textwidth]{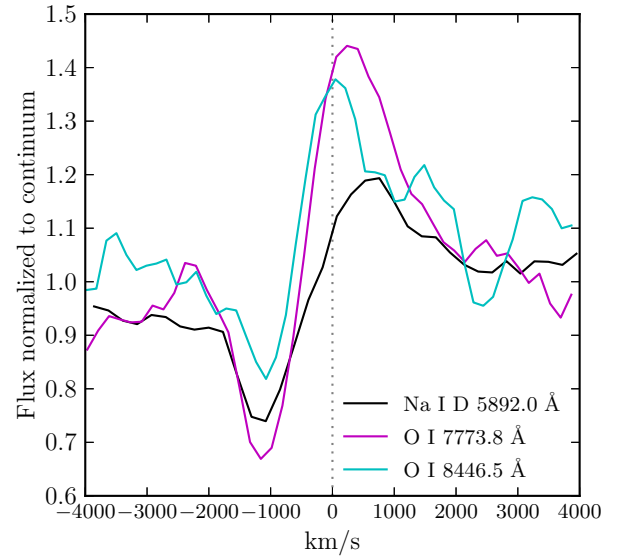}
\caption{\ion{Na}{1}~D and \ion{O}{1} P~Cygni profiles from the 2010 February 7.30 spectrum, taken $1.03~\textrm{d}$ after maximum light. The P~Cygni profiles trace the ejecta velocities from the nova explosion, since at early phases the ejecta is optically thick and approximates the true expansion velocity. \ion{Na}{1}~D and \ion{O}{1} are doublet and triplet lines, respectively, and are centered relative to the weighted means of the NIST atomic line database relative strengths. The line flux is normalized to continuum.}
\label{fig:pcygnis}
\end{center}
\end{figure}

\begin{figure*}[htb]
\begin{center}
\includegraphics[width=0.8\textwidth]{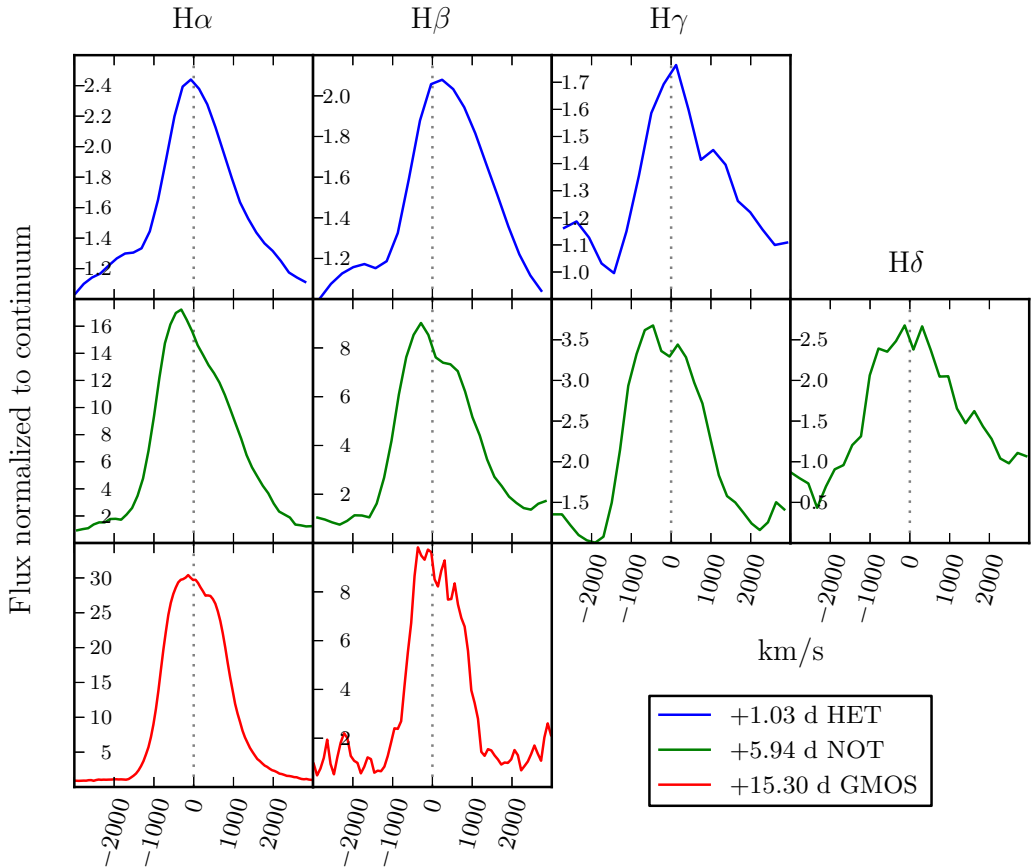}
\caption{Hydrogen Balmer line profiles for \obj. The profiles become narrower and more box-like with time, and the FWHM decreases with time. The line profiles are scaled relative to the continuum flux.}
\label{fig:balmer_profiles}
\end{center}
\end{figure*}

\begin{figure}
\begin{center}
\includegraphics[width=0.48\textwidth]{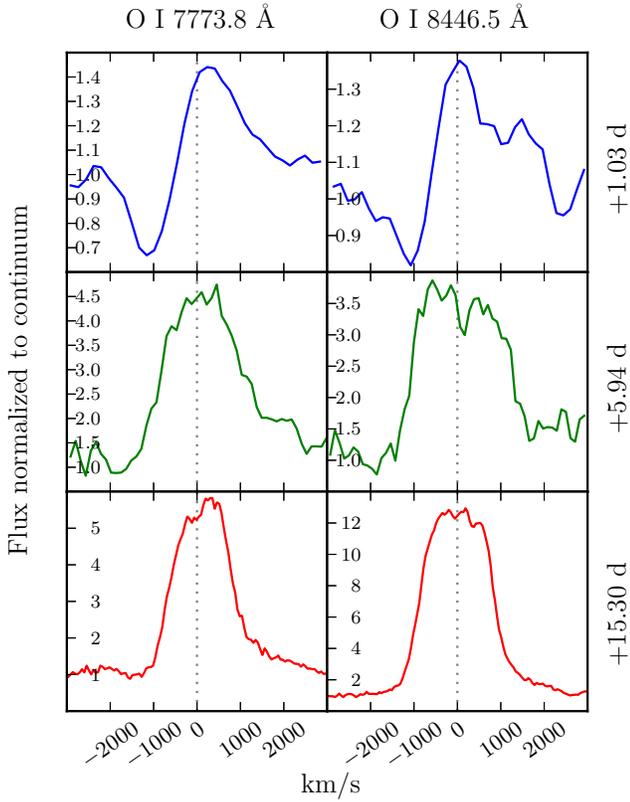}
\caption{\ion{O}{1} $\lambda7774$ and $\lambda8446$ line profiles.  Initially, these are P~Cygni line profiles characteristic of an optically thick expanding envelope that develop into a flat-topped profile, characteristic of an optically thin expanding shell. Flux is normalized to the continuum.}
\label{fig:OI_8446}
\end{center}
\end{figure}

\begin{figure}
\begin{center}
\includegraphics[width=0.48\textwidth]{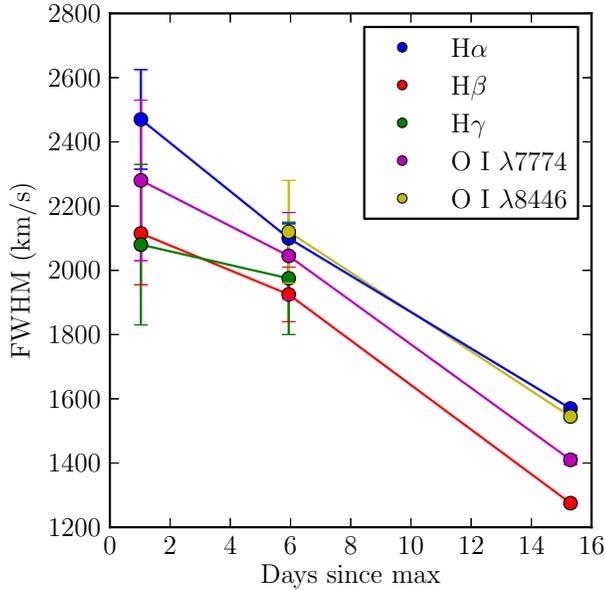}
\caption{All velocities plotted are FWHM except the $t=0$~d \ion{O}{1} points, which are derived from the P~Cygni profiles. For compatibility with this plot, line species with P Cygni profiles (\ion{O}{1}) are plotted as twice the velocity of the blue shifted absorption component. The FWHM of the lines decreases with time.}
\label{fig:line_velocities}
\end{center}
\end{figure}

Strong \pc\ profiles are clearly seen in the \ion{Na}{1} D $\lambda5892$ and \ion{O}{1} $\lambda7774$ lines (Figure~\ref{fig:pcygnis}). The presence of these profiles in the 2010 \HET\ spectrum (Figure~\ref{fig:overall_spectra}) are characteristic of spectra of novae at maximum light \citep{war03}, therefore with this additional information to the initial rise in $r^\prime$ band, it is likely that the light curve (Figure~\ref{fig:lightcurve}) captures the maximum light of \obj. We take an average of the velocities of the P~Cygni lines of \ion{Na}{1}~D and \ion{O}{1} (Figure~\ref{fig:pcygnis}) and the widths of the Balmer series and \ion{O}{1} (Figures~\ref{fig:balmer_profiles}~\&~\ref{fig:OI_8446}) to derive an expansion velocity of $\approx 1100 \kms$ (Figure~\ref{fig:line_velocities}).

When the envelope is initially optically thick the radiation is ionization bounded and neutral and low-ionization emission lines are formed. As the nova evolves, the ionizing radiation becomes progressively harder as the photosphere recedes to the surface of the hot white dwarf and higher ionization states are seen. \citet{w90} determines that for electron number densities $N_e \gtrsim 10^9 {\rm cm}^{-3}$ the nova envelope is optically thick, while forbidden lines will appear once $N_e \lesssim 10^7 {\rm cm}^{-3}$.

By 5.94 d after peak (2010 \NOT), the P~Cygni profiles become pure emission while the Balmer and \ion{Fe}{2} emission lines are still clearly visible. The wider wavelength range of the NOT spectrum reveals \ion{Ca}{2} H \& K emission lines and additional Balmer series lines continuing until the Balmer break. The most significant change is the increase in strength of the \ion{O}{1} $\lambda7773$ and $\lambda8446$ lines and the appearance of the forbidden lines of [\ion{O}{1}] $\lambda5577$, $6300$ and $6363$.

\begin{figure*}[htb]
\begin{center}
  \includegraphics[width=0.7\textwidth]{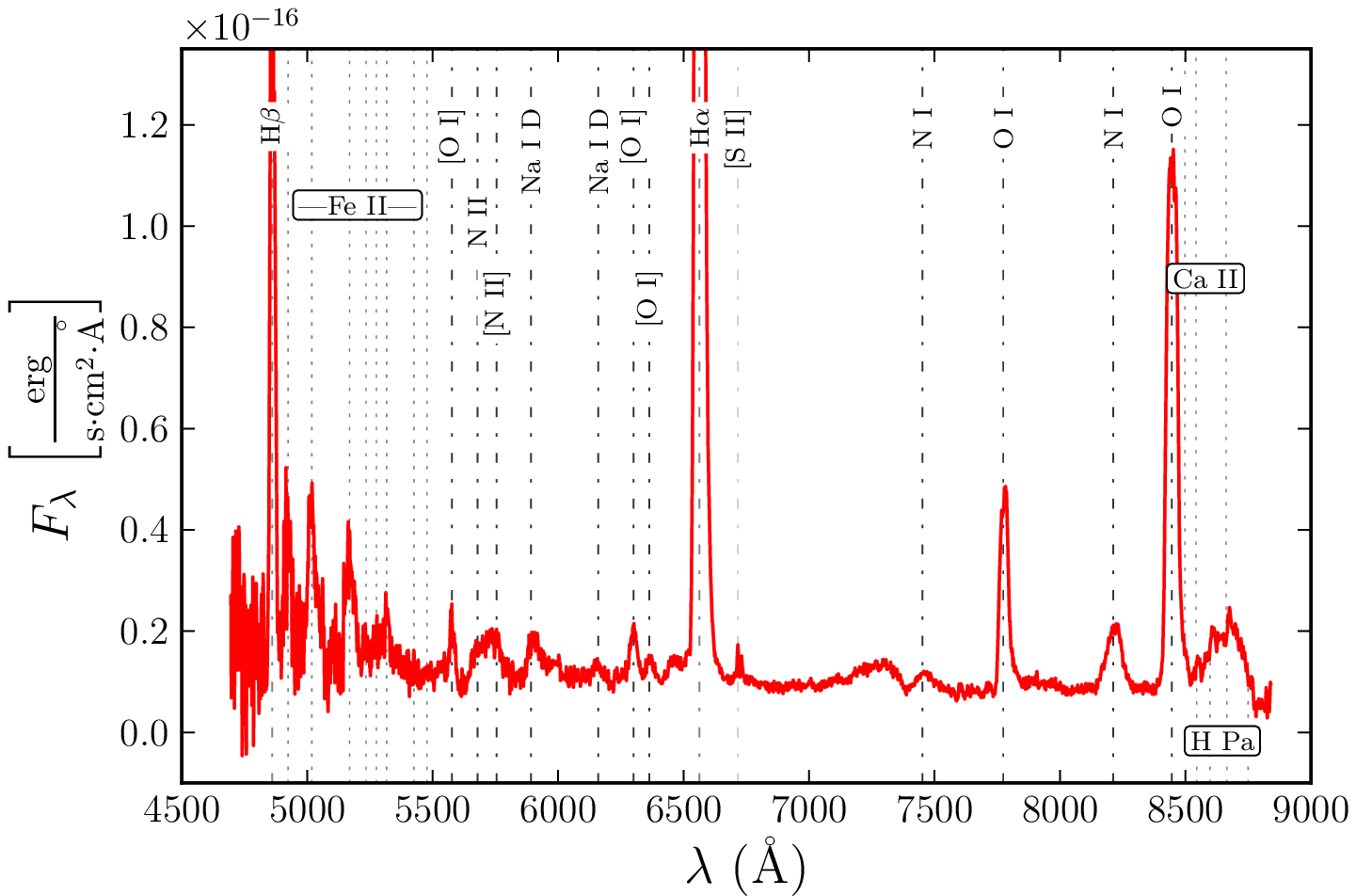}
\caption{Late time (15.3 d after maximum) spectrum of \obj. The strong presence of \ion{O}{1} and the first appearance of [\ion{O}{1}] emission suggests that \obj\ is entering the nebular stage. Host galaxy lines (such as [\ion{S}{2}]) are labeled in grey, while multiplet species are labeled with dotted lines.}
\label{fig:GMOS_spectrum}
\end{center}
\end{figure*}

By 15.30~d after peak the \ion{Fe}{2} emission lines have mostly faded and the Balmer series dominates in emission (Figure~\ref{fig:GMOS_spectrum}). Throughout all spectral epochs, the Balmer lines are the strongest emission lines (Figure~\ref{fig:balmer_profiles}), evolving from a FWHM of $2200 \kms$ at 1.03~d after maximum light to $1600\kms$ at 15.30~d after maximum light. Initially, the Balmer profiles show an asymmetric structure, but then evolve to become narrower and more symmetric. The O~I~$\lambda8446$~\AA\ line developed a flat-topped profile (Figure~\ref{fig:OI_8446}) characteristic of an optically thin expanding spherical shell at a velocity of $815\kms$. There are also faint forbidden lines of [\ion{O}{1}], signaling the entrance into the nebular phase of classical nova spectral evolution. The late time spectrum is characteristic of a nova shell, showing strong Balmer lines, \ion{O}{1}, and signs of [\ion{O}{1}]~$\lambda6300$~\AA\, which signals the transition from the permitted state to the forbidden state. 

\citet{sls+11} argues that, based only upon an early spectrum from Keck/LRIS on 2010 February 7th UT (2 d after maximum light), the spectra and light curve are very similar to a LBV, (SN)\,2000ch \citep{wvh+04,pbt+10}. However, upon examination of extant nova spectra, a surprisingly close match to \obj\ is found with L91 (Figure~\ref{fig:LMC_comp_all}). The prominent \ion{Fe}{2} and \ion{O}{1} emission visible in the late time spectrum additionally suggests that \obj\ is not an LBV because most known LBVs do not show this behavior \citep{sls+11}.

\citet{wil92} devises an optical spectroscopic classification system that types novae by their strongest non-Balmer emission lines, typically \ion{Fe}{2} or a combination of He and N, called He/N. They find that He/N novae preferentially have shorter $t_2$, higher expansion velocities, and coronal lines, while \ion{Fe}{2} novae evolve to a forbidden line spectrum with lower ionization species. The He/N spectrum is formed in a discrete shell ejected during the explosive thermonuclear runaway while the \ion{Fe}{2} spectrum is formed in a continuous wind driven by the radiation from the residual burning of material on the surface of the white dwarf.  \citet{wil92} explains that manifestation of the spectrum is dependent on which mechanism dominates in a two-component model. The spectral evolution of \obj\ and its evolution clearly identifies it as a member of the \ion{Fe}{2} spectral class. 

We observe velocity evolution in the emission line profiles of \obj\ (Figure~\ref{fig:line_velocities}), with the Balmer and \ion{O}{1} $\lambda7774$ and $\lambda8446$ profiles exhibiting a jagged shape at early times, and then becoming smoother and narrower.  For \ion{Fe}{2} novae, this is the result of a photosphere formed in a wind with velocity homologously increasing outward and mass loss rate decreasing with time.  The decreasing density pushes the region of line formation steadily inward towards the surface of the white dwarf where flow velocities are lower.

\citet{sno+11} find that the presence of [\ion{Fe}{10}] 6375 requires a hot photoionization source, and thus correlates well with Super Soft X-ray emission. That this line is not visible in the nebular spectra complements the non-detection of X-ray emission from \obj.

\section{\obj\ as a Fast and Extremely Luminous Classical Nova}

\subsection{Comparison to Nova LMC 1991 and M31N-2007-11d}

Although roughly a thousand classical novae have been discovered, only a few luminous events are known due to their rapid decline and intrinsic rarity. Two other luminous \ion{Fe}{2} type novae have been studied extensively: L91 \citep{del91,sss+01,wph94} and M31N-2007-11d \citep{srq+09}, hereafter M31N.  L91 was an exceedingly bright and fast \ion{Fe}{2} type nova in the Large Magellanic Cloud, so luminous that it was initially heralded as a prototype for a class of super-bright novae \citep{del91}. M31N was discovered during a spectroscopic survey of novae in M31 by \citet{sdh+11}.

\begin{figure}
\begin{center}
\includegraphics[width=0.48\textwidth]{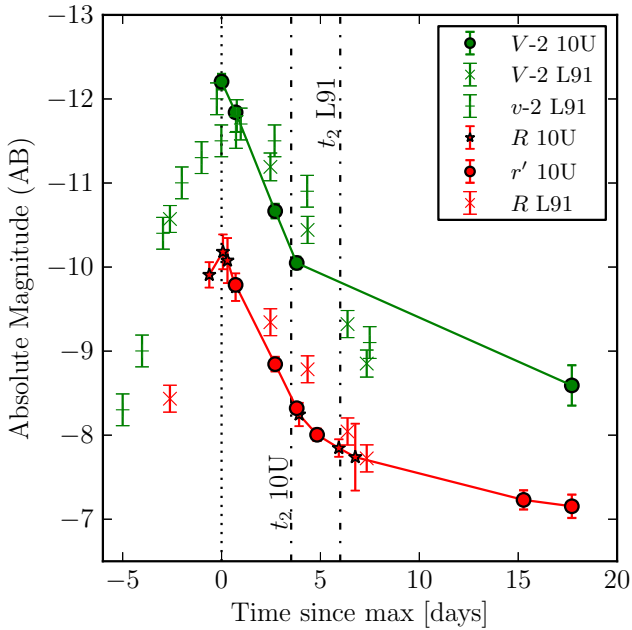}
\caption{Comparison of \obj\ (10U) to L91. ``$v$'' represents visual magnitude, and $v$-band and $V$-band measurements are offset by two magnitudes for clarity. Both novae exhibit similar peak absolute brightnesses and decline rates. The rise time of L91 (and M31N) is exceptionally long, while the rise time of 10U is unconstrained.}
\label{fig:lightcurve_LMC_comp}
\end{center}
\end{figure}

The light curves of L91 and M31N are similar to \obj\ (a comparison between \obj\ and L91 is shown in Figure~\ref{fig:lightcurve_LMC_comp}). L91 was discovered 5 days before maximum light and M31N $\gtrsim 4$~d before maximum. The rise to maximum of L91 is among the longest for novae on record, with a peak of $M_v = -10.0$ mag. The light curve of L91 shown here is drawn from the photometry published in the circulars \citep{sss+91,gil91,glc+91,lmh+91,dlm+91}.  \citet{srq+09} set a lower limit of 4~d on the rise time for M31N from quiescence to a maximum light of $M_V \simeq -9.5$ mag. Both novae declined rapidly from maximum light with $t_2 = 6\pm1$~d (L91: \citealt{sss+01}) and $t_2 = 9.5$~d (M31N: \citealt{srq+09}). By comparison, \obj\ has $t_2 = 3.5 \pm 0.3$~d and the rise time is unconstrained.

\begin{figure*}[htb]
\begin{center}
\includegraphics{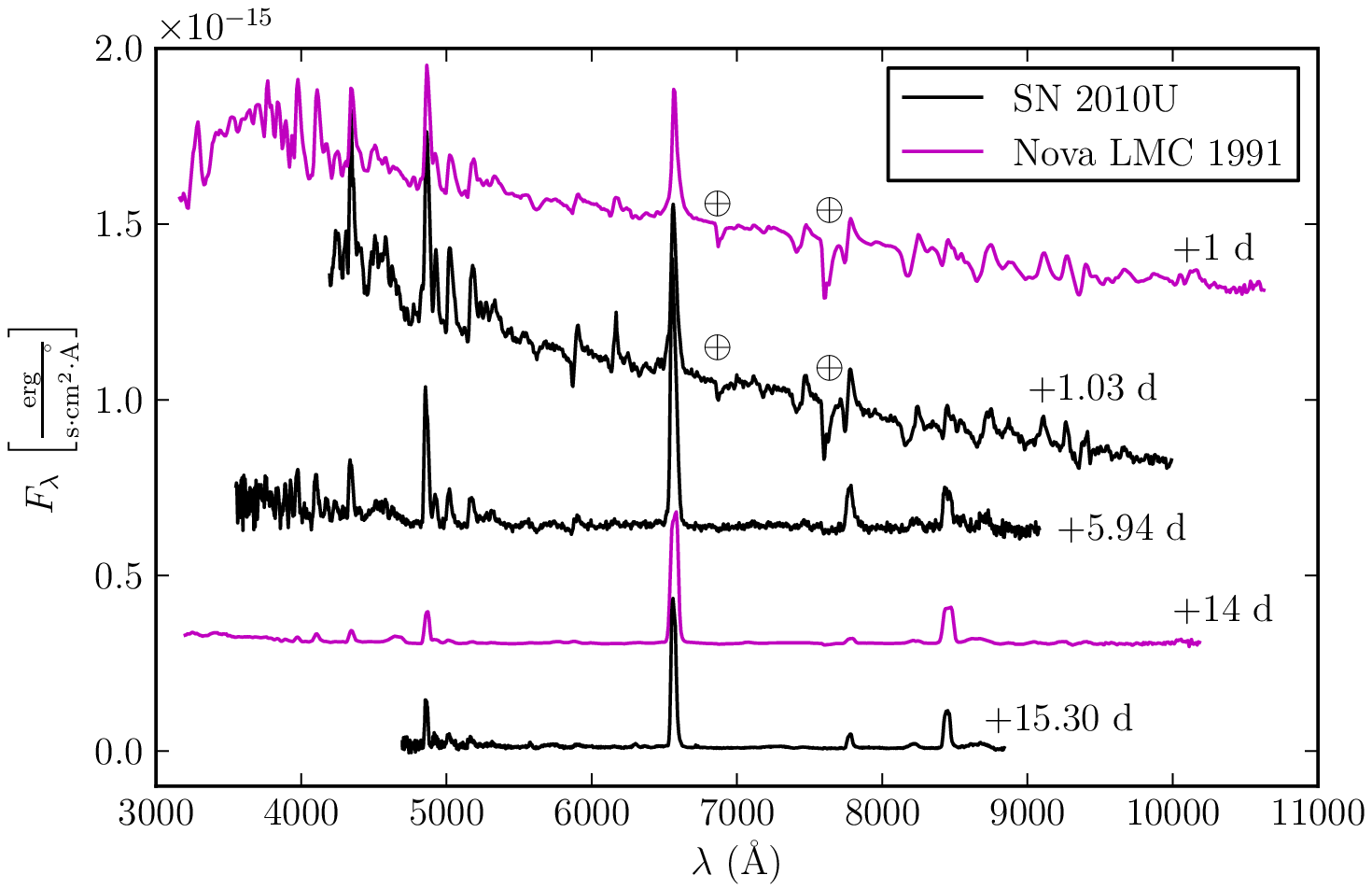}
\caption{Spectroscopic comparison to L91. Both novae exhibit strong \ion{Fe}{2} lines, hydrogen Balmer profiles with similar widths, P~Cygni profiles at early times, and similar timescales for spectral evolution.}
\label{fig:LMC_comp_all}
\end{center}
\end{figure*}

Spectroscopically, L91 and M31N are remarkably similar to \obj--they are all clearly \ion{Fe}{2} type novae. L91 and M31N have slightly lower expansion velocities with H$\alpha$~FWHM of $\simeq 1880\kms$ and $\simeq 1550\kms$, respectively, while \obj\ has $\simeq 2230\kms$. At early times both L91 and M31N show strong P~Cygni absorption profiles. L91 clearly mirrors the temporal and spectral evolution of \obj\ (Figure~\ref{fig:LMC_comp_all}).

L91 is one of the best studied novae of the modern era. UV spectra from the \emph{IUE} satellite revealed strong \ion{Fe}{2} absorption which would be reradiated as emission in the optical \citep{sss+01}. \citet{sss+01} construct a model atmosphere of L91 using ${\tt PHOENIX}$ and ${\tt CLOUDY}$ to obtain abundance estimates of the outburst and find that L91 was enriched in CNO elements and originated from a carbon oxygen (CO) white dwarf.

Although there is no late time spectroscopy of \obj, the spectrum 15.94~d after maximum light already shows evidence of forbidden oxygen, with no evidence for any neon.  This, combined with the presence of carbon and oxygen and the similarity of spectra to L91, suggests that \obj\ also had a carbon-oxygen WD progenitor. 

However, efforts to identify WD progenitor types are confounded by the possibility that an enriched envelope could exist on top of a CO white dwarf, or that an ONeMg nova may or may not have a dredge-up event that would enrich the spectrum, producing a wide range of observable spectra. Whether or not there is a direct mapping between the manifestation of the spectrum and the composition of the underlying WD is still an open question \citep{pk98,mas11}, although with detailed UV and X-ray spectral observations capturing the entirety of the nova outburst, such as in the case of L91 \citep{sss+01}, it may be possible to tell.

\subsection{MMRD and FWHM Relationships}

Studies of novae have revealed a correlation between peak absolute magnitude $M_V$ and decline rate $t_2$, termed the maximum magnitude versus rate of decline relationship (MMRD). The shorter $t_2$, the more intrinsically luminous the nova explosion \citep{dl95,dd00,sdh+11}.

\begin{figure}
\begin{center}
\includegraphics[width=0.48\textwidth]{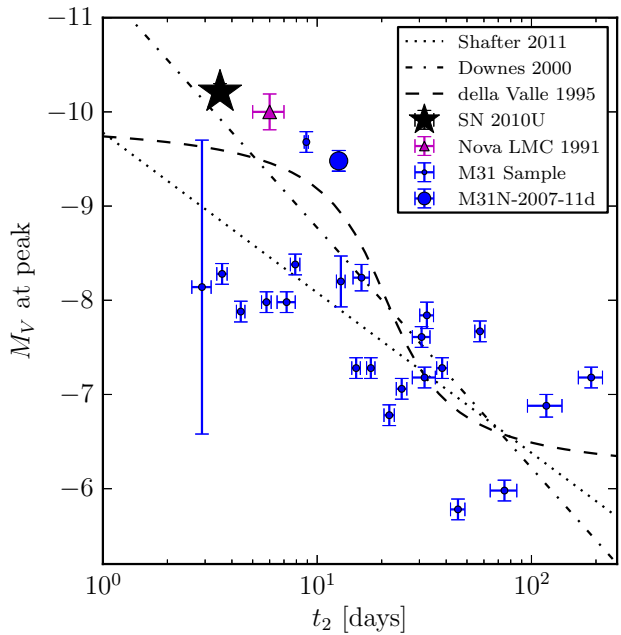}
\caption{Maximum magnitude vs. rate of decline (MMRD) relationship created from data from the \citet{sdh+11} survey with luminous novae and historical MMRD relationships overplotted.}
\label{fig:MMRD_V}
\end{center}
\end{figure}

\begin{figure*}
\begin{center}
  \includegraphics[width=0.9\textwidth]{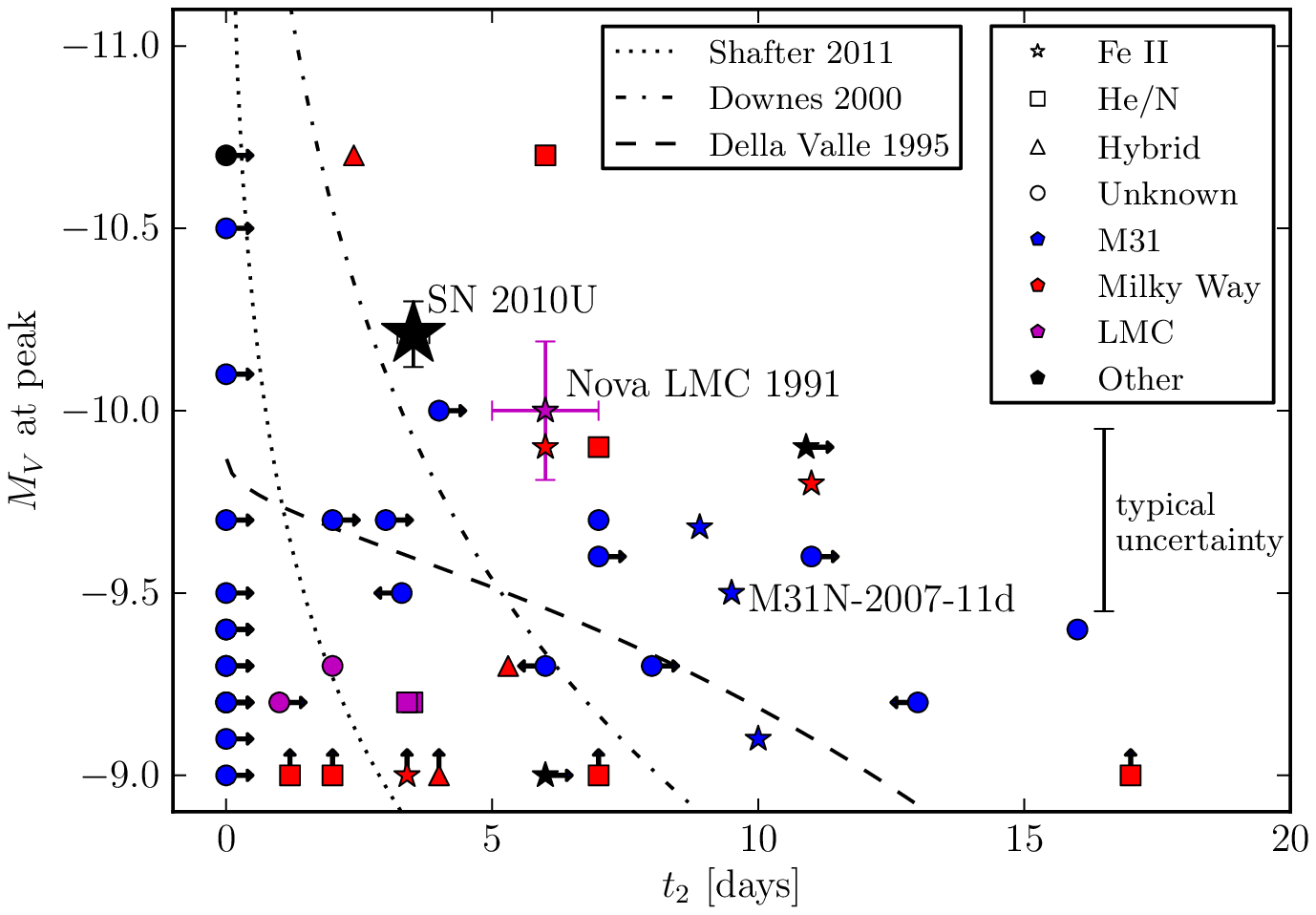}
\caption{MMRD relationships \citep{dl95,dd00,sdh+11} plotted over the most luminous novae on record \citet{srq+09,kck+10,sdh+11}. Upper or lower limits on $t_2$ are shown with arrows, and novae for which no $t_2$ measurement exists are plotted at $t_2 = 0$. The colors denote location and the symbols denote spectral type. All magnitudes are $V$-band except for the black points other than \obj, which are measured in $g^\prime$ \citep{kck+10}. The typical uncertainty is shown, which is largely an uncertainty of whether the nova was caught at peak magnitude.}
\label{fig:extreme_spec_MMRD}
\end{center}
\end{figure*}

\citet{sdh+11} executed an extensive multi-year study of novae in M31, discovering and spectroscopically classifying 46 novae, bringing the total of spectroscopically classified novae in M31 to 91. They derive a MMRD for M31 novae and compare to other historical samples of novae (Figure~\ref{fig:MMRD_V}) \citep{dl95,dd00,sdh+11}. There is a substantial amount of scatter in this relation. The extreme quadrant of the MMRD at high luminosities and shortest $t_2$ is shown in Figure~\ref{fig:extreme_spec_MMRD} with the most luminous novae known to date compiled by \citet{srq+09} and updated with recent discoveries by \citet{kck+10} and \citet{sdh+11}. Although in Figure~\ref{fig:extreme_spec_MMRD} there appear to be a comparable number of luminous He/N novae and \ion{Fe}{2} novae, because He/N novae are rarer they are in fact preferentially brighter and faster than \ion{Fe}{2} novae \citep{wil92, sdh+11, sdb+12}. For example, \citet{sdh+11} find in their M31 survey that three of their four fastest declining novae are He/N type, although by number He/N novae comprise only $\sim 20$\% of all novae.

Despite claims that there might be a ``super-bright'' class of novae \citep{del91}, \citet{srq+09} find no evidence for a distinct population. However, \obj\ is indeed a very luminous nova: compared to the 883 novae on record compiled by \citet{srq+09}, only 4 novae are brighter, two of which are not spectroscopically confirmed \citep{cfn+87,kck+10}. At the extreme end of the luminosity distribution there is large scatter from observational uncertainties such as intrinsic extinction and uncertainty in the capture of maximum light as well as uncertainties from intrinsic variability in the novae explosion due to variation in WD mass, accretion rate and metallicity \citep{srq+09}. 

\begin{figure}
\begin{center}
\includegraphics[width=0.48\textwidth]{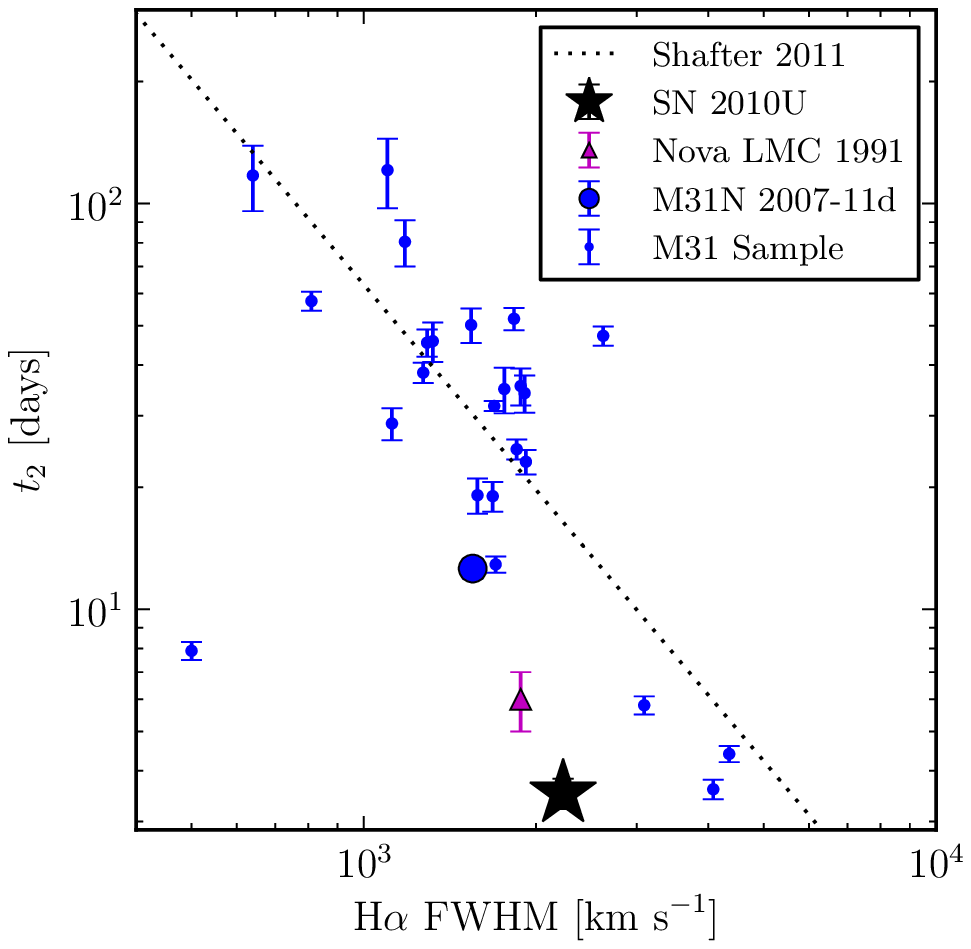}
\caption{H$\alpha$ width compared to decline speed (measured by $t_2$) of the nova outburst for the \citet{sdh+11} sample of novae from M31, with L91 and \obj\ also plotted. \obj\ and L91 are at slightly slower velocities than the mean, but still consistent with the trend, given the large scatter.  Interestingly, the three bright \ion{Fe}{2} novae are all faster declining than the mean.}
\label{fig:vel_t2}
\end{center}
\end{figure}

Surveys also find that novae with faster expansion velocities have a faster decline from maximum light (Figure~\ref{fig:vel_t2}) \citep{mcl60,sdh+11}. The scatter in Figure~\ref{fig:vel_t2} is likely due to the time dependence of velocities, which depends on how soon after maximum light the spectra was obtained. Such a $t_2$ vs. H$\alpha$~FWHM relationship is a natural outcome if He/N novae are the fastest declining and the most violent, having the highest ejecta velocities. \obj\ is a fast declining nova with a moderate ejection velocity.

\subsection{Nova Populations and Progenitor}

Since the spectroscopic classification scheme of \citet{whp+91} was devised and the MMRD predicted that more massive WDs produce more violent explosions (see \S~\ref{sec:super_edd}) \citep{dl95}, it has been an outstanding question whether the spectral type of a nova correlates with the properties of the underlying stellar population. In general, the nova population follows the galaxy light \citep{sdh+11,sdb+12}. Surveys of multiple galaxies of different morphologies aim to determine whether novae properties such as peak brightness, $t_2$, and spectral type correlate with the underlying stellar population.

In the Milky Way, \citet{dl98} find that He/N novae are concentrated at the Galactic plane and are fast and bright, while \ion{Fe}{2} novae are concentrated in the bulge and thick disk of the galaxy and are slow and dim. Because younger stellar populations have on average more massive white dwarfs, disk novae should be more extreme; therefore they claim that He/N novae are associated with a younger stellar population and that the \ion{Fe}{2} novae are associated with an older stellar population. 

In M31, \citet{sdh+11} find conflicting results with no compelling evidence that spectroscopic class depends on location within the galaxy. However, they did find that the spatial distribution He/N novae is slightly more extended than that of \ion{Fe}{2} novae and that the spatial distribution of faster (lower $t_2$) novae is slightly more extended than that of slower novae. 

In M33, however, \citet{sdb+12} find that five of eight novae are of He/N or \ion{Fe}{2}b hybrid novae, while only two are definitively of the \ion{Fe}{2} class. Hybrid novae initially have broad \ion{Fe}{2} lines which then later are replaced by He/N lines. Interestingly, the opposite evolution from He/N to \ion{Fe}{2} has never been observed. They speculate that this statistically significant difference in the fraction of He/N to \ion{Fe}{2} Galactic and M31 novae could be a result of the underlying population of M33, which is a bulgeless galaxy, and therefore would be expected to be dominated by a disk population.

To address whether or not the most luminous \ion{Fe}{2} novae are associated with a particular stellar population, it is useful to investigate the associated stellar populations of L91, M31N, and \obj. \citet{sa02} examine the region surrounding L91 and find that there are three clusters within $\sim130$~pc with ages less than the young age of $10^{7.5}$~yr, and that the location of L91 is close to another fast nova, LMC 1977\#2. Because the LMC is a bulgeless galaxy, we would expect it to be dominated by fast declining novae, and indeed it possesses a fast declining and fast ejecta population of novae \citep{dd93}. However, that the luminous \ion{Fe}{2} nova L91 specifically came from a young stellar population is in tension with the prediction that \ion{Fe}{2} novae are associated with older stellar populations.

The location of M31N-2007-11d is at large galactocentric radius from the center of M31 and perhaps a member of the disk, although another luminous but less-studied \ion{Fe}{2} nova M31N-2009-09b is close to the center of the galaxy. However, the large inclination of M31 makes it difficult to determine whether or not M31N-2009-09b is actually in the bulge or might be within the disk and projected in front of the bulge. 

\citet{hpr+10} use pre-explosion archival \emph{Hubble Space Telescope} images with WFC3 F814W and WFPC2 F555W and F814W to investigate the progenitor of \obj\ and the associated stellar population. At the location of the nova, there is a photometric limit of $M_V \approx -3.2$ mag, which puts an upper limit on the mass of the progenitor system and its companion of $3-5~M_\odot$. The stars within 100~pc of the location of the nova are dim and red, suggesting association with an evolved population. The spatial location of any younger main sequence stars is distinct from the location of the nova, suggesting it is not associated with a massive star population, although it could be an evolved and obscured lower mass AGB star \citep{hpr+10}. If \obj\ originated from an evolved population, this would follow the emerging trend that \ion{Fe}{2} novae come from lower mass, older populations such as the bulge of M31. 

Based upon the theory of the MMRD (see \S~\ref{sec:super_edd}), the high intrinsic luminosity and fast temporal evolution of \obj\ signal that the progenitor was a high mass WD. The evolutionary channels of how the binary system could reach its pre-outburst state depend on when the WD was born. The WD could be born massive from a massive star, or alternatively the WD could have formed from a less massive star and accreted material. If the WD was born massive, there is a higher chance that it would be an ONeMg WD. Some claim that ``neon novae,'' which exhibit strong [\ion{Ne}{3}] and [\ion{Ne}{5}] lines during the forbidden phase, and thus have a high neon abundance, originate from ONeMg WDs \citep{mas11}, although it is unclear whether or not these lines could be produced by an enriched surface layer and be uncorrelated with WD composition \citep{pk98}. The nova models of \citet{yps+05} are able to produce the full range of observed nova characteristics ($M_V$, $t_2$, $v_{\rm ejecta}$) using only CO WD progenitors, however with ONeMg WD progenitors they were able to produce ejecta envelopes that were enriched in neon by $\sim 10^4$ times compared to CO WDs. 

Surveys have established that most luminous and fast novae are He/N novae, however, \obj, L91, and M31N are all \ion{Fe}{2} novae. \citet{srq+09} speculate that what may set these novae apart is their long rise time compared to the general nova population, which reaches maximum light in less than 3 days. \citet{sss+01} hypothesize that the long rise time is indicative of a large amount of ejected mass, such that the photons take a long time to diffuse and escape. For L91, \citet{sss+01} found $M_{\rm ej} \sim 3 \times 10^{-4} M_\odot$ with a progenitor of a high mass, cool WD with a low metallicity envelope. \citet{srq+09} speculate that a long rise time may be related to the formation of the \ion{Fe}{2} spectrum, which is formed in an optically thick wind driven by residual burning on the surface of the WD. 

\subsection{Nova Physics and Super Eddington Luminosity}
\label{sec:super_edd}

\citet{dl95} proposed that the MMRD is primarily a function of the mass of the WD progenitor. If the WD is more massive, the surface layers will be more degenerate and allow a more intense but also more rapid expulsion of material.  Recent studies suggest that the outburst properties additionally depend sensitively upon other parameters of the progenitor system such as the temperature of the isothermal core and the accretion rate of material onto the surface of the WD. \citet{tb05} calculate the ignition mass of the accreted material on the surface of the WD, $M_{\rm ign}$, and its dependence on the mass transfer rate  $\dot{M}$ to the WD, and the WD mass. The temperature of the WD core, $T_c$, also influences $M_{\rm ign}$; however, \citet{tb04} find that $T_c$ is set uniquely by $\dot{M}$, and therefore $M_{\rm ign}$ is primarily a function of only two parameters. A lower $\dot{M} \lesssim 10^{-10} M_\odot\; \text{yr}^{-1}$ leads to a lower $T_c$, which then increases the $M_{\rm ign}$ needed to start the nova eruption. Alternatively, a high $\dot{M} \gtrsim 10^{-9} M_\odot\; \text{yr}^{-1}$ will trigger the thermonuclear runaway earlier; at extremely high $\dot{M}$, stable hydrogen burning can occur \citep{tb04}.

\citet{yps+05} show that very low accretion rates can produce the most extreme nova explosions, which are characterized by super-Eddington luminosities at maximum, large ejecta velocities, fast optical decline, and if the WD is of moderate mass, large ejecta masses. The extreme luminosities and rapid photometric declines of L91, M31N, and \obj\ suggest that these novae all originated from massive WDs. \citet{sss+01} claim that L91 had a $\gtrsim 1.2 M_\odot$ WD. Comparing the outburst characteristics ($t_2$, bolometric luminosity, and ejecta velocity) of \obj\ to the grid of theoretical predictions made by \citet{yps+05}, we find that only the models with massive ($M \approx 1.25 M_\odot$) and cool ($T = 3 \times 10^7$~K) white dwarfs accreting at a very low rate ($\dot{M} \lesssim 10^{-11} M_\odot {\rm yr}^{-1}$) are able to reproduce these parameters. 

\citet{sss+01} speculate that the large luminosity of L91, which would otherwise be inconsistent with the large ejecta mass, could be the result of a traveling shock wave through colliding ejecta shells. As described in \citet{wil92}, there is a discrete low density and high velocity shell and an optically thick wind which is powered by nuclear burning of residual material on the surface of the white dwarf. He/N spectra are dominated by the discrete component, while \ion{Fe}{2} novae are dominated by the wind component. A more massive ejecta shell would be more likely to have residual material which to burn on the surface of the WD, and may explain why the massive ejecta of L91 and likely massive ejecta of \obj\ result in \ion{Fe}{2} novae.

The rise times of novae are generally longer than that of those predicted by spherically symmetric models, which suggests that time might be needed for the local thermonuclear runaway to proceed over the surface of the white dwarf \citep{war03}, and may result in an asymmetric outburst. \citet{whp+91} observe transient absorption features in high resolution line profiles of Balmer and \ion{He}{1} lines in other novae that are likely due to discrete absorption components, such as a small cloud of high density passing in front of the continuum source, suggesting the outburst is inhomogeneous. Clumpy ejecta would increase the effective Eddington limit and allow the nova outburst to sustain apparently super-Eddington luminosities for a period of time \citep{sha01}. Given that nova ejecta are inhomogeneous and quickly evolving with time, it is likely that the super-Eddington novae L91 and \obj\ sustained their remarkable luminosities through a porous photosphere or asymmetric explosion. 

Although He/N novae are preferentially brighter than \ion{Fe}{2} novae, there are several very luminous \ion{Fe}{2} novae, in particular L91, M31N, and \obj. Although the MMRD and population studies suggest that \ion{Fe}{2} novae should predominately come from older stellar populations which have on average smaller mass WDs, this mapping between white dwarf and spectral type must not necessarily be direct. When a binary stellar system will evolve to a configuration that can produce a nova outburst is not a simple function of the mass or age of the dwarf, but is dependent upon the orbital parameters and stellar evolution of the binary system itself, for example, when the orbit might decay or the donor might evolve to fill its Roche lobe and begin mass transfer. Bright \ion{Fe}{2} novae like L91, M31N, and \obj\ may represent unique binary systems with cool, high mass CO white dwarfs accreting material from their companions at a very low rate. 

\section{Conclusions}

\obj\ was a luminous (peak $M_V = -10.2 \pm 0.1$ mag) and fast declining ($t_2 = 3.5 \pm 0.3$ d) classical nova in the galaxy NGC\,4214.  Optical spectroscopy revealed that it was an \ion{Fe}{2} type nova with strong hydrogen Balmer emission and expansion velocities of order $\approx 1100\kms$.  P~Cygni spectral line profiles in spectra taken near maximum light indicate that the emission was optically thick and early photometry indicates that the optical emission was approximately thermal blackbody with $T \approx 8000$~K.  As the nova faded, the spectrum evolved to a nebular state dominated by emission lines and [\ion{O}{1}] emission began to appear.

Our conclusions are the following:

\begin{enumerate}
  
\item \obj\ was a fast and luminous nova, among the top 0.5\% brightest of all historical outbursts and the third brightest nova for which spectroscopic information exists.  It is remarkably similar to both Nova LMC 1991 (L91) and M31N-2007-11d (M31N) in photometry and spectra.
  
\item \obj\ is a \ion{Fe}{2} type nova. \ion{Fe}{2} novae are characteristically dimmer and slower to decline (longer $t_2$) than He/N. The existence of bright and fast \ion{Fe}{2} novae like \obj, L91, and M31N are interesting outliers in trends which aim to correlate spectral type with outburst properties.
  
\item \obj\ reached super-Eddington luminosities during the peak of its outburst. Most novae are sub-Eddington, however L91 was also super-Eddington for an extended period of time while it ejected a large amount of mass. It is likely that the \ion{Fe}{2} spectrum, which is formed in an optically thick wind, is related to high mass loss.
  
\item Massive and luminous nova outbursts like \obj\ probe a unique set of progenitor parameters, and point to an extreme region of parameter space with low accretion rate, high white dwarf mass, and low white dwarf core temperature. The extreme luminosity region of the MMRD is poorly constrained and subject to high scatter.
  
\item That \obj\ likely originated from a CO WD associated with an evolved stellar population is interesting in the context of the debate of nova populations and the manifestation of nova spectral type. Trends in the Milky Way, the LMC, and M33 suggest that more luminous novae of the He/N type originate from young stellar populations where average white dwarf mass is higher. However, L91, M31N, and \obj\ all are extremely luminous \ion{Fe}{2} novae that are likely from massive CO white dwarfs. Various paths of binary evolution can influence when these systems will enter a configuration that would generate nova outbursts.

\end{enumerate}

Upcoming wide-field transient surveys like LSST will discover optical transients in ever greater numbers.  In particular, the high cadence and deep optical limits of the survey will reveal many classical novae, which have traditionally been difficult to study because of moderate luminosities and fast decline from maximum light.  Understanding the extreme quadrant of high luminosity and rapid optical decline for classical novae is paramount for maximizing the scientific return of large photometric surveys, for which spectroscopic resources will not be available to confirm every discovery. Accurately characterizing the intermediate luminosity phase space now will be paramount to understanding the wealth of data from future transient surveys.

\section*{Acknowledgments}

\acknowledgments IC thanks Greg Schwarz, Robert Williams, Maxwell Moe, Alica Soderberg, Warren Brown, James Moran, and Rosanne Di Stefano for valuable conversations about nova physics, and Wen-fai Fong for helpful comments on this manuscript.  We graciously acknowledge the amateur astronomers who contributed critical discovery images: K.~Itagaki, J.~Brimacombe, T.~Yusa, and J.~Nicolas.  IC is supported by the NSF Graduate Research Fellowship Grant. JCW's group at UT Austin is supported by NSF Grant AST 11-09881.  JV is supported by Hungarian OTKA grant K76816.  This research was completed through the use of many facilities.  It is based on observations made with the Nordic Optical Telescope, operated on the island of La Palma jointly by Denmark, Finland, Iceland, Norway, and Sweden, in the Spanish Observatorio del Roque de los Muchachos of the Instituto de Astrofisica de Canarias. The data presented here were obtained in part with ALFOSC, which is provided by the Instituto de Astrofisica de Andalucia (IAA) under a joint agreement with the University of Copenhagen and NOTSA.  The Liverpool Telescope is operated on the island of La Palma by Liverpool John Moores University in the Spanish Observatorio del Roque de los Muchachos of the Instituto de Astrofisica de Canarias with financial support from the UK Science and Technology Facilities Council. This research is based on observations obtained at the Gemini Observatory under program GN-2010A-Q-30 and principal investigator Edo Berger. Gemini Observatory is operated by the Association of Universities for Research in Astronomy, Inc., under a cooperative agreement with the NSF on behalf of the Gemini partnership: the National Science Foundation (United States), the Science and Technology Facilities Council (United Kingdom), the National Research Council (Canada), CONICYT (Chile), the Australian Research Council (Australia), Minist\'{e}rio da Ci\^{e}ncia e Tecnologia (Brazil) and Ministerio de Ciencia, Tecnolog\'{i}a e Innovaci\'{o}n Productiva (Argentina). We acknowledge the use of public data from the Swift data archive. This research has made use of data obtained through the High Energy Astrophysics Science Archive Research Center Online Service, provided by the NASA/Goddard Space Flight Center. This research has made use of the XRT Data Analysis Software (XRTDAS) developed under the responsibility of the ASI Science Data Center (ASDC), Italy. The NIST Atomic Spectra Database was used: Ralchenko, Yu., Kramida, A.E., Reader, J., and NIST ASD Team (2011). NIST Atomic Spectra Database (ver. 4.1.0), [Online]. Available: http://physics.nist.gov/asd3 [2011, August 1]. National Institute of Standards and Technology, Gaithersburg, MD.

\bibliographystyle{apj}

\begin{thebibliography}{62}
\expandafter\ifx\csname natexlab\endcsname\relax\def\natexlab#1{#1}\fi

\bibitem[{{Balman} {et~al.}(1998){Balman}, {Krautter}, \& {Oegelman}}]{bko98}
{Balman}, S., {Krautter}, J., \& {Oegelman}, H. 1998, \apj, 499, 395

\bibitem[{{Berger} {et~al.}(2009){Berger}, {Soderberg}, {Chevalier},
  {Fransson}, {Foley}, {Leonard}, {Debes}, {Diamond-Stanic}, {Dupree}, {Ivans},
  {Simmerer}, {Thompson}, \& {Tremonti}}]{bsc+09}
{Berger}, E., {et~al.} 2009, \apj, 699, 1850

\bibitem[{{Bode} \& {Evans}(2008)}]{be08}
{Bode}, M.~F., \& {Evans}, A. 2008, {Classical Novae}, ed. {Bode, M.~F.~\&
  Evans, A.}

\bibitem[{{Botticella} {et~al.}(2009){Botticella}, {Pastorello}, {Smartt},
  {Meikle}, {Benetti}, {Kotak}, {Cappellaro}, {Crockett}, {Mattila}, {Sereno},
  {Patat}, {Tsvetkov}, {van Loon}, {Abraham}, {Agnoletto}, {Arbour}, {Benn},
  {di Rico}, {Elias-Rosa}, {Gorshanov}, {Harutyunyan}, {Hunter}, {Lorenzi},
  {Keenan}, {Maguire}, {Mendez}, {Mobberley}, {Navasardyan}, {Ries},
  {Stanishev}, {Taubenberger}, {Trundle}, {Turatto}, \& {Volkov}}]{bps+09}
{Botticella}, M.~T., {et~al.} 2009, \mnras, 398, 1041

\bibitem[{{Burrows} {et~al.}(2005){Burrows}, {Hill}, {Nousek}, {Kennea},
  {Wells}, {Osborne}, {Abbey}, {Beardmore}, {Mukerjee}, {Short}, {Chincarini},
  {Campana}, {Citterio}, {Moretti}, {Pagani}, {Tagliaferri}, {Giommi},
  {Capalbi}, {Tamburelli}, {Angelini}, {Cusumano}, {Br{\"a}uninger}, {Burkert},
  \& {Hartner}}]{bhn+05}
{Burrows}, D.~N., {et~al.} 2005, \ssr, 120, 165

\bibitem[{{Ciardullo} {et~al.}(1987){Ciardullo}, {Ford}, {Neill}, {Jacoby}, \&
  {Shafter}}]{cfn+87}
{Ciardullo}, R., {Ford}, H.~C., {Neill}, J.~D., {Jacoby}, G.~H., \& {Shafter},
  A.~W. 1987, \apj, 318, 520

\bibitem[{{Dalcanton} {et~al.}(2009){Dalcanton}, {Williams}, {Seth}, {Dolphin},
  {Holtzman}, {Rosema}, {Skillman}, {Cole}, {Girardi}, {Gogarten},
  {Karachentsev}, {Olsen}, {Weisz}, {Christensen}, {Freeman}, {Gilbert},
  {Gallart}, {Harris}, {Hodge}, {de Jong}, {Karachentseva}, {Mateo}, {Stetson},
  {Tavarez}, {Zaritsky}, {Governato}, \& {Quinn}}]{dws+09}
{Dalcanton}, J.~J., {et~al.} 2009, \apjs, 183, 67

\bibitem[{{Darnley} {et~al.}(2006){Darnley}, {Bode}, {Kerins}, {Newsam}, {An},
  {Baillon}, {Belokurov}, {Calchi Novati}, {Carr}, {Cr{\'e}z{\'e}}, {Evans},
  {Giraud-H{\'e}raud}, {Gould}, {Hewett}, {Jetzer}, {Kaplan},
  {Paulin-Henriksson}, {Smartt}, {Tsapras}, \& {Weston}}]{dbk+06}
{Darnley}, M.~J., {et~al.} 2006, \mnras, 369, 257

\bibitem[{{Della Valle}(1991)}]{del91}
{Della Valle}, M. 1991, \aap, 252, L9

\bibitem[{{Della Valle} \& {Duerbeck}(1993)}]{dd93}
{Della Valle}, M., \& {Duerbeck}, H.~W. 1993, \aap, 271, 175

\bibitem[{{Della Valle} {et~al.}(1991){Della Valle}, {Leisy}, {McNaught},
  {Savage}, {Hartley}, {Hughes}, \& {Garradd}}]{dlm+91}
{Della Valle}, M., {Leisy}, P., {McNaught}, R.~H., {Savage}, A., {Hartley}, M.,
  {Hughes}, S.~M., \& {Garradd}, G. 1991, \iaucirc, 5260, 1

\bibitem[{{Della Valle} \& {Livio}(1998)}]{dl98}
{Della Valle}, M., \& {Livio}, M. 1998, \apj, 506, 818

\bibitem[{{Djupvik} \& {Andersen}(2010)}]{da10}
{Djupvik}, A.~A., \& {Andersen}, J. 2010, in Highlights of Spanish Astrophysics
  V, ed. {J.~M.~Diego, L.~J.~Goicoechea, J.~I.~Gonz{\'a}lez-Serrano, \&
  J.~Gorgas}, 211--+

\bibitem[{{Downes} \& {Duerbeck}(2000)}]{dd00}
{Downes}, R.~A., \& {Duerbeck}, H.~W. 2000, \aj, 120, 2007

\bibitem[{{Filippenko}(1997)}]{fil97}
{Filippenko}, A.~V. 1997, \araa, 35, 309

\bibitem[{{Gehrels} {et~al.}(2004){Gehrels}, {Chincarini}, {Giommi}, {Mason},
  {Nousek}, {Wells}, {White}, {Barthelmy}, {Burrows}, {Cominsky}, {Hurley},
  {Marshall}, {M{\'e}sz{\'a}ros}, {Roming}, {Angelini}, {Barbier}, {Belloni},
  {Campana}, {Caraveo}, {Chester}, {Citterio}, {Cline}, {Cropper}, {Cummings},
  {Dean}, {Feigelson}, {Fenimore}, {Frail}, {Fruchter}, {Garmire}, {Gendreau},
  {Ghisellini}, {Greiner}, {Hill}, {Hunsberger}, {Krimm}, {Kulkarni}, {Kumar},
  {Lebrun}, {Lloyd-Ronning}, {Markwardt}, {Mattson}, {Mushotzky}, {Norris},
  {Osborne}, {Paczynski}, {Palmer}, {Park}, {Parsons}, {Paul}, {Rees},
  {Reynolds}, {Rhoads}, {Sasseen}, {Schaefer}, {Short}, {Smale}, {Smith},
  {Stella}, {Tagliaferri}, {Takahashi}, {Tashiro}, {Townsley}, {Tueller},
  {Turner}, {Vietri}, {Voges}, {Ward}, {Willingale}, {Zerbi}, \&
  {Zhang}}]{gcg+04}
{Gehrels}, N., {et~al.} 2004, \apj, 611, 1005

\bibitem[{{Gilmore}(1991)}]{gil91}
{Gilmore}, A.~C. 1991, \iaucirc, 5253, 3

\bibitem[{{Gilmore} {et~al.}(1991){Gilmore}, {Liller}, {Cooper}, \&
  {Overbeek}}]{glc+91}
{Gilmore}, A.~C., {Liller}, W., {Cooper}, T., \& {Overbeek}, D. 1991, \iaucirc,
  5250, 1

\bibitem[{{G{\"u}ver} \& {{\"O}zel}(2009)}]{go09}
{G{\"u}ver}, T., \& {{\"O}zel}, F. 2009, \mnras, 400, 2050

\bibitem[{{Hachisu} \& {Kato}(2010)}]{hk10}
{Hachisu}, I., \& {Kato}, M. 2010, \apj, 709, 680

\bibitem[{{Hill} {et~al.}(1998){Hill}, {Nicklas}, {MacQueen}, {Tejada}, {Cobos
  Duenas}, \& {Mitsch}}]{hnm+98}
{Hill}, G.~J., {Nicklas}, H.~E., {MacQueen}, P.~J., {Tejada}, C., {Cobos
  Duenas}, F.~J., \& {Mitsch}, W. 1998, in Presented at the Society of
  Photo-Optical Instrumentation Engineers (SPIE) Conference, Vol. 3355, Society
  of Photo-Optical Instrumentation Engineers (SPIE) Conference Series, ed.
  {S.~D'Odorico}, 375--386

\bibitem[{{Hook} {et~al.}(2004){Hook}, {J{\o}rgensen}, {Allington-Smith},
  {Davies}, {Metcalfe}, {Murowinski}, \& {Crampton}}]{had+04}
{Hook}, I.~M., {J{\o}rgensen}, I., {Allington-Smith}, J.~R., {Davies}, R.~L.,
  {Metcalfe}, N., {Murowinski}, R.~G., \& {Crampton}, D. 2004, \pasp, 116, 425

\bibitem[{{Humphreys} \& {Davidson}(1994)}]{hd94}
{Humphreys}, R.~M., \& {Davidson}, K. 1994, \pasp, 106, 1025

\bibitem[{{Humphreys} {et~al.}(2010){Humphreys}, {Prieto}, {Rosenfield},
  {Helton}, {Kochanek}, {Stanek}, {Khan}, {Szczygiel}, {Mogren}, {Fesen},
  {Milisavljevic}, {Williams}, {Murphy}, {Dalcanton}, \& {Gilbert}}]{hpr+10}
{Humphreys}, R.~M., {et~al.} 2010, \apjl, 718, L43

\bibitem[{Jones {et~al.}(2001)Jones, Oliphant, Peterson, {et~al.}}]{jop+01}
Jones, E., Oliphant, T., Peterson, P., {et~al.} 2001, {SciPy}: Open source
  scientific tools for {Python}

\bibitem[{{Kasliwal} {et~al.}(2010){Kasliwal}, {Cenko}, {Kulkarni}, {Ofek},
  {Quimby}, {Rau}, {Caltech}, {Berkeley}, \& {Garching}}]{kck+10}
{Kasliwal}, M.~M., {et~al.} 2010, ArXiv e-prints

\bibitem[{{Landolt}(1992)}]{lan92}
{Landolt}, A.~U. 1992, \aj, 104, 340

\bibitem[{{Liller} {et~al.}(1991){Liller}, {McNaught}, {Hughes}, {Hartley},
  {Camilleri}, \& {Garradd}}]{lmh+91}
{Liller}, W., {McNaught}, R.~H., {Hughes}, S.~M., {Hartley}, M., {Camilleri},
  P., \& {Garradd}, G. 1991, \iaucirc, 5244, 1

\bibitem[{{Mason}(2011)}]{mas11}
{Mason}, E. 2011, \aap, 532, L11

\bibitem[{{McLaughlin}(1960)}]{mcl60}
{McLaughlin}, D.~B. 1960, in Stellar Atmospheres, ed. J.~L. {Greenstein}, 585

\bibitem[{{Nakano} \& {Kadota}(2010)}]{nk10}
{Nakano}, S., \& {Kadota}, K. 2010, Central Bureau Electronic Telegrams, 2161,
  1

\bibitem[{{Pastorello} {et~al.}(2010){Pastorello}, {Botticella}, {Trundle},
  {Taubenberger}, {Mattila}, {Kankare}, {Elias-Rosa}, {Benetti}, {Duszanowicz},
  {Hermansson}, {Beckman}, {Bufano}, {Fraser}, {Harutyunyan}, {Navasardyan},
  {Smartt}, {van Dyk}, {Vink}, \& {Wagner}}]{pbt+10}
{Pastorello}, A., {et~al.} 2010, \mnras, 408, 181

\bibitem[{{Poole} {et~al.}(2008){Poole}, {Breeveld}, {Page}, {Landsman},
  {Holland}, {Roming}, {Kuin}, {Brown}, {Gronwall}, {Hunsberger}, {Koch},
  {Mason}, {Schady}, {vanden Berk}, {Blustin}, {Boyd}, {Broos}, {Carter},
  {Chester}, {Cucchiara}, {Hancock}, {Huckle}, {Immler}, {Ivanushkina},
  {Kennedy}, {Marshall}, {Morgan}, {Pandey}, {de Pasquale}, {Smith}, \&
  {Still}}]{pbp+08}
{Poole}, T.~S., {et~al.} 2008, \mnras, 383, 627

\bibitem[{{Prialnik} \& {Kovetz}(1998)}]{pk98}
{Prialnik}, D., \& {Kovetz}, A. 1998, in Astronomical Society of the Pacific
  Conference Series, Vol. 137, Wild Stars in the Old West, ed. S.~{Howell},
  E.~{Kuulkers}, \& C.~{Woodward}, 376

\bibitem[{{Prieto} {et~al.}(2008){Prieto}, {Kistler}, {Thompson}, {Y{\"u}ksel},
  {Kochanek}, {Stanek}, {Beacom}, {Martini}, {Pasquali}, \&
  {Bechtold}}]{pkt+08}
{Prieto}, J.~L., {et~al.} 2008, \apjl, 681, L9

\bibitem[{{Ramsey} {et~al.}(1998){Ramsey}, {Adams}, {Barnes}, {Booth},
  {Cornell}, {Fowler}, {Gaffney}, {Glaspey}, {Good}, {Hill}, {Kelton},
  {Krabbendam}, {Long}, {MacQueen}, {Ray}, {Ricklefs}, {Sage}, {Sebring},
  {Spiesman}, \& {Steiner}}]{rab+98}
{Ramsey}, L.~W., {et~al.} 1998, in Presented at the Society of Photo-Optical
  Instrumentation Engineers (SPIE) Conference, Vol. 3352, Society of
  Photo-Optical Instrumentation Engineers (SPIE) Conference Series, ed.
  {L.~M.~Stepp}, 34--42

\bibitem[{{Roming} {et~al.}(2005){Roming}, {Kennedy}, {Mason}, {Nousek}, {Ahr},
  {Bingham}, {Broos}, {Carter}, {Hancock}, {Huckle}, {Hunsberger}, {Kawakami},
  {Killough}, {Koch}, {McLelland}, {Smith}, {Smith}, {Soto}, {Boyd},
  {Breeveld}, {Holland}, {Ivanushkina}, {Pryzby}, {Still}, \& {Stock}}]{rkm+05}
{Roming}, P.~W.~A., {et~al.} 2005, \ssr, 120, 95

\bibitem[{{Schlegel} {et~al.}(1998){Schlegel}, {Finkbeiner}, \&
  {Davis}}]{sfd98}
{Schlegel}, D.~J., {Finkbeiner}, D.~P., \& {Davis}, M. 1998, \apj, 500, 525

\bibitem[{{Schwarz} {et~al.}(2001){Schwarz}, {Shore}, {Starrfield},
  {Hauschildt}, {Della Valle}, \& {Baron}}]{sss+01}
{Schwarz}, G.~J., {Shore}, S.~N., {Starrfield}, S., {Hauschildt}, P.~H., {Della
  Valle}, M., \& {Baron}, E. 2001, \mnras, 320, 103

\bibitem[{{Schwarz} {et~al.}(2011){Schwarz}, {Ness}, {Osborne}, {Page},
  {Evans}, {Beardmore}, {Walter}, {Helton}, {Woodward}, {Bode}, {Starrfield},
  \& {Drake}}]{sno+11}
{Schwarz}, G.~J., {et~al.} 2011, \apjs, 197, 31

\bibitem[{{Shafter} {et~al.}(2012){Shafter}, {Darnley}, {Bode}, \&
  {Ciardullo}}]{sdb+12}
{Shafter}, A.~W., {Darnley}, M.~J., {Bode}, M.~F., \& {Ciardullo}, R. 2012,
  ArXiv e-prints

\bibitem[{{Shafter} {et~al.}(2009){Shafter}, {Rau}, {Quimby}, {Kasliwal},
  {Bode}, {Darnley}, \& {Misselt}}]{srq+09}
{Shafter}, A.~W., {Rau}, A., {Quimby}, R.~M., {Kasliwal}, M.~M., {Bode}, M.~F.,
  {Darnley}, M.~J., \& {Misselt}, K.~A. 2009, \apj, 690, 1148

\bibitem[{{Shafter} {et~al.}(2011){Shafter}, {Darnley}, {Hornoch},
  {Filippenko}, {Bode}, {Ciardullo}, {Misselt}, {Hounsell}, {Chornock}, \&
  {Matheson}}]{sdh+11}
{Shafter}, A.~W., {et~al.} 2011, \apj, 734, 12

\bibitem[{{Shaviv}(2001)}]{sha01}
{Shaviv}, N.~J. 2001, \mnras, 326, 126

\bibitem[{{Shore} {et~al.}(1991){Shore}, {Starrfield}, {Sonneborn}, {Gilmore},
  {Liller}, {Jones}, \& {Pearce}}]{sss+91}
{Shore}, S.~N., {Starrfield}, S.~G., {Sonneborn}, G., {Gilmore}, A.~C.,
  {Liller}, W., {Jones}, A., \& {Pearce}, A. 1991, \iaucirc, 5257, 1

\bibitem[{{Smith} {et~al.}(2011){Smith}, {Li}, {Silverman}, {Ganeshalingam}, \&
  {Filippenko}}]{sls+11}
{Smith}, N., {Li}, W., {Silverman}, J.~M., {Ganeshalingam}, M., \&
  {Filippenko}, A.~V. 2011, \mnras, 415, 773

\bibitem[{{Steele} {et~al.}(2004){Steele}, {Smith}, {Rees}, {Baker}, {Bates},
  {Bode}, {Bowman}, {Carter}, {Etherton}, {Ford}, {Fraser}, {Gomboc}, {Lett},
  {Mansfield}, {Marchant}, {Medrano-Cerda}, {Mottram}, {Raback}, {Scott},
  {Tomlinson}, \& {Zamanov}}]{ssr+04}
{Steele}, I.~A., {et~al.} 2004, in Presented at the Society of Photo-Optical
  Instrumentation Engineers (SPIE) Conference, Vol. 5489, Society of
  Photo-Optical Instrumentation Engineers (SPIE) Conference Series, ed.
  {J.~M.~Oschmann Jr.}, 679--692

\bibitem[{{Subramaniam} \& {Anupama}(2002)}]{sa02}
{Subramaniam}, A., \& {Anupama}, G.~C. 2002, \aap, 390, 449

\bibitem[{{Szczygie{\l}} {et~al.}(2012){Szczygie{\l}}, {Prieto}, {Kochanek},
  {Stanek}, {Thompson}, {Beacom}, {Garnavich}, \& {Woodward}}]{spk+12}
{Szczygie{\l}}, D.~M., {Prieto}, J.~L., {Kochanek}, C.~S., {Stanek}, K.~Z.,
  {Thompson}, T.~A., {Beacom}, J.~F., {Garnavich}, P.~M., \& {Woodward}, C.~E.
  2012, \apj, 750, 77

\bibitem[{{Thompson} {et~al.}(2009){Thompson}, {Prieto}, {Stanek}, {Kistler},
  {Beacom}, \& {Kochanek}}]{tps+09}
{Thompson}, T.~A., {Prieto}, J.~L., {Stanek}, K.~Z., {Kistler}, M.~D.,
  {Beacom}, J.~F., \& {Kochanek}, C.~S. 2009, \apj, 705, 1364

\bibitem[{{Townsley} \& {Bildsten}(2004)}]{tb04}
{Townsley}, D.~M., \& {Bildsten}, L. 2004, \apj, 600, 390

\bibitem[{{Townsley} \& {Bildsten}(2005)}]{tb05}
---. 2005, \apj, 628, 395

\bibitem[{{Valenti} {et~al.}(2011){Valenti}, {Fraser}, {Benetti}, {Pignata},
  {Sollerman}, {Inserra}, {Cappellaro}, {Pastorello}, {Smartt}, {Ergon},
  {Botticella}, {Brimacombe}, {Bufano}, {Crockett}, {Eder}, {Fugazza},
  {Haislip}, {Hamuy}, {Harutyunyan}, {Ivarsen}, {Kankare}, {Kotak}, {LaCluyze},
  {Magill}, {Mattila}, {Maza}, {Mazzali}, {Reichart}, {Taubenberger},
  {Turatto}, \& {Zampieri.}}]{vfb+11}
{Valenti}, S., {et~al.} 2011, ArXiv e-prints

\bibitem[{Valle \& Livio(1995)}]{dl95}
Valle, M.~D., \& Livio, M. 1995, \apj, 452, 704

\bibitem[{{van den Bergh} \& {Younger}(1987)}]{vy87}
{van den Bergh}, S., \& {Younger}, P.~F. 1987, \aaps, 70, 125

\bibitem[{{Wagner} {et~al.}(2004){Wagner}, {Vrba}, {Henden}, {Canzian},
  {Luginbuhl}, {Filippenko}, {Chornock}, {Li}, {Coil}, {Schmidt}, {Smith},
  {Starrfield}, {Klose}, {Tich{\'a}}, {Tich{\'y}}, {Gorosabel}, {Hudec}, \&
  {Simon}}]{wvh+04}
{Wagner}, R.~M., {et~al.} 2004, \pasp, 116, 326

\bibitem[{{Warner}(2003)}]{war03}
{Warner}, B. 2003, {Cataclysmic Variable Stars}, ed. {Warner, B.}

\bibitem[{{Williams}(1990)}]{w90}
{Williams}, R.~E. 1990, in Lecture Notes in Physics, Berlin Springer Verlag,
  Vol. 369, IAU Colloq. 122: Physics of Classical Novae, ed. A.~{Cassatella} \&
  R.~{Viotti}, 213

\bibitem[{{Williams}(1992)}]{wil92}
{Williams}, R.~E. 1992, \aj, 104, 725

\bibitem[{{Williams} {et~al.}(1991){Williams}, {Hamuy}, {Phillips},
  {Heathcote}, {Wells}, \& {Navarrete}}]{whp+91}
{Williams}, R.~E., {Hamuy}, M., {Phillips}, M.~M., {Heathcote}, S.~R., {Wells},
  L., \& {Navarrete}, M. 1991, \apj, 376, 721

\bibitem[{{Williams} {et~al.}(1994){Williams}, {Phillips}, \& {Hamuy}}]{wph94}
{Williams}, R.~E., {Phillips}, M.~M., \& {Hamuy}, M. 1994, \apjs, 90, 297

\bibitem[{{Yaron} {et~al.}(2005){Yaron}, {Prialnik}, {Shara}, \&
  {Kovetz}}]{yps+05}
{Yaron}, O., {Prialnik}, D., {Shara}, M.~M., \& {Kovetz}, A. 2005, \apj, 623,
  398

\end{thebibliography}

\section{Tables}

\begin{deluxetable}{rcr@{$\pm$}lr@{$\pm$}lr@{$\pm$}ll}
\tablecaption{Photometry of \obj\ \label{table:phot}}
\tablewidth{0pt}
\tablehead{
\colhead{2010 UT} & \colhead{Filter} & \multicolumn{2}{c}{$m\pm\sigma_m$ (Vega)}
& \multicolumn{2}{c}{$m\pm\sigma_m$ (AB, de-ext)} &
\multicolumn{2}{l}{$f_\nu$ ($\mu$Jy$\pm\sigma$)} & \colhead{Observer/Telescope}
}
\startdata
Jan 24.74 & $R$ & \multicolumn{2}{l}{$>18.80$} & \multicolumn{2}{l}{$>18.92$}
& \multicolumn{2}{l}{$< 98$} & Itagaki unfiltered\\
Feb 5.65 & $R$ & 17.38 & 0.15 & 17.50 & 0.15 & 362 & 86 & Itagaki unfiltered\\
Feb 6.27 & $V$ & 17.28 & 0.09 & 17.20 & 0.09 & 477 & 66 & Brimacombe
unfiltered\\
Feb 6.34 & $R$ & 17.10 & 0.20 & 17.23 & 0.21 & 466 & 153 & Yusa unfiltered\\
Feb 6.57 & $R$ & 17.21 & 0.27 & 17.33 & 0.27 & 424 & 180 & Itagaki unfiltered\\
Feb 6.98 & $R$ & 17.52 & 0.16 & 17.65 & 0.16 & 317 & 82 & Nicolas\\
Feb 6.98 & $B$ & 18.09 & 0.04 & 17.88 & 0.04 & 255 & 17 & LT\\
Feb 6.98 & $V$ & 17.64 & 0.05 & 17.57 & 0.05 & 341 & 27 & LT\\
Feb 6.98 & $r^\prime$ & 17.54 & 0.04 & 17.62 & 0.04 & 325 & 20 & LT\\
Feb 6.98 & $i^\prime$ & 17.42 & 0.04 & 17.73 & 0.04 & 294 & 20 & LT\\
Feb 8.98 & $B$ & 18.90 & 0.12 & 18.69 & 0.12 & 121 & 23 & LT\\
Feb 8.98 & $V$ & 18.82 & 0.08 & 18.74 & 0.08 & 115 & 15 & LT\\
Feb 8.98 & $r^\prime$ & 18.48 & 0.08 & 18.56 & 0.08 & 136 & 18 & LT\\
Feb 8.98 & $i^\prime$ & 18.42 & 0.07 & 18.73 & 0.07 & 117 & 13 & LT\\
Feb 10.07 & $B$ & 19.59 & 0.08 & 19.38 & 0.08 & 64 & 8 & LT\\
Feb 10.07 & $V$ & 19.43 & 0.07 & 19.36 & 0.07 & 65 & 7 & LT\\
Feb 10.07 & $r^\prime$ & 19.01 & 0.04 & 19.09 & 0.04 & 84 & 5 & LT\\
Feb 10.07 & $i^\prime$ & 18.96 & 0.06 & 19.27 & 0.06 & 71 & 7 & LT\\
Feb 10.17 & $i^\prime$ & 18.99 & 0.04 & 19.30 & 0.04 & 69 & 4 & LT\\
Feb 10.19 & $R$ & 19.04 & 0.14 & 19.16 & 0.14 & 78 & 17 & NOT unfiltered\\
Feb 11.10 & $r^\prime$ & 19.32 & 0.07 & 19.40 & 0.07 & 63 & 7 & LT\\
Feb 12.20 & $R$ & 19.44 & 0.10 & 19.56 & 0.10 & 54 & 9 & NOT unfiltered\\
Feb 13.03 & $R$ & 19.55 & 0.40 & 19.67 & 0.40 & 49 & 31 & Nicolas\\
Feb 21.54 & $g^\prime$ &  \multicolumn{2}{l}{\nodata} & 21.15 & 0.05 & 13 & 1 &
GMOS\\
Feb 21.54 & $r^\prime$ &  \multicolumn{2}{l}{\nodata} & 20.18 & 0.11 & 31 & 5
& GMOS\\
Feb 21.54 & $i^\prime$ & \multicolumn{2}{l}{\nodata}  & 20.35 & 0.08 & 26 & 3
& GMOS\\
Feb 23.98 & $B$ & \multicolumn{2}{l}{$>20.89$} & \multicolumn{2}{l}{$>20.68$} &
\multicolumn{2}{l}{$<19$}& LT\\
Feb 23.98 & $V$ & 20.89 & 0.24 & 20.82 & 0.24 & 17 & 7 & LT\\
Feb 23.98 & $r^\prime$ & 20.17 & 0.14 & 20.26 & 0.14 & 29 & 6 & LT\\
Feb 23.98 & $i^\prime$ & 20.32 & 0.18 & 20.63 & 0.18 & 20 & 6 & LT\\
\enddata
\tablecomments{
Unfiltered images were calibrated to Landolt $R$-band using field stars from the Landolt catalog. One exception to this was the unfiltered image from J.~Brimacombe on 2010 February 6.27~UT, which was calibrated to standard $V$-band (see \S~\ref{sec:phot}). Raw photometry is reported in the Vega system, uncorrected for Galactic or host-galaxy extinction. To put all measurements on a uniform scale, the Vega measurements were converted to AB magnitudes using the formula $m_{\rm x, AB} = m_{x, VEGA} + \Delta m_{\rm x}$, where $\Delta m_{\rm x}$ is the offset derived from ${\tt pysynphot}$: $\Delta m_B = -0.115$, $\Delta m_V =0.000$, $\Delta m_R = 0.183$, $\Delta m_{r^\prime} = 0.142$ and $\Delta m_{i^\prime} = 0.356$. These AB magnitude and Janksy values in the table are corrected for galactic reddening of $E(B-V) = 0.022$, while the Vega values are uncorrected. Observations that were originally in AB were not converted back to Vega.}
\end{deluxetable}

\begin{deluxetable}{ccccc}
\tablecaption{\emph{Swift XRT} observations of \obj \label{table:swift_xrt}}
\tablehead{
\colhead{UT Date} & \colhead{Exposure (s)} &
\colhead{Counts $({\rm ct\;s}^{-1})$} & \colhead{Flux $({\rm erg\;s}^{-1})$ $(kT=60\;{\rm eV})$} & \colhead{Flux $({\rm erg\;s}^{-1})$ $(kT=5\;{\rm keV})$}
}
\startdata
2007 March 26.50 UT & 5624 & $<$6.67E-03 & $<1.9\times 10^{-13}$ & $<1.1\times10^{-12}$\\
2010 March 3.82 UT & 1903 & $<$9.08E-03 & $<2.6\times 10^{-13}$ & $<1.5\times 10^{-12}$\\
\enddata
\tablecomments{Energy band $0.3 - 10.0$~keV. All upper limits are $3\sigma$.}
\end{deluxetable}

\begin{deluxetable}{cccrc}
\tablecaption{\label{table:swift_uvot}\emph{Swift UVOT} photometry of \obj
}
\tablehead{
\colhead{UT Date (March 2010)} & \colhead{Filter} & \colhead{$\lambda$
(\AA)} & \colhead{Exposure (s)} & \colhead{$m$ (AB)}
}
\startdata
3.97 & UVM2 & 2246 & 465 & $>20.63$ \\
3.82 & UVW1 & 2600 & 594 & $>20.88$ \\
3.82 & $U$ & 3465 & 141 & $>19.82$ \\
3.82 & $B$ & 4392 & 141 & $>19.15$ \\
3.97 & $V$ & 5468 & 60 & $>17.88$ \\
\enddata
\tablecomments{All observations taken 2010 March 3.82 UT. All upper limits are $3\sigma$.}
\end{deluxetable}

\begin{deluxetable}{rllcccccc}
\tablecolumns{7}
\tablewidth{0pc}
\tablecaption{\obj\ Spectroscopic Observations \label{table:spectroscopy}}
\tablehead{
\colhead{$\Delta t$\tablenotemark{a} (d)} & \colhead{2010 UT} &
\colhead{Telescope}
& \colhead{Instrument} &
\colhead{$\lambda$ (\AA)} & \colhead{Resolution (\AA)} & \colhead{Exposure (s)} &
\colhead{Airmass} & \colhead{Slit Width (arcsec)} }
\startdata
+1.03 & Feb 7.30 & HET & LRS & 4200-10000 & 15 & 437 & 1.23 & 2 \\
+5.94 & Feb 12.21 & NOT & ALFOSC & 3200-9100 & 15 & $3 \times 900$ & 1.05 &
1\\
+15.30 & Feb 21.57 & GN & GMOS & 4700 - 8840 & 8 & $2 \times 1200$ & 1.07 &
1\\
\enddata
\tablecomments{Maximum light was on 2010 February \maxdate.}
\tablenotetext{a}{$\Delta t = t - t_{\rm max}$, where $t_{\rm max} =
$~2010 Feb~6.27 UT.}
\end{deluxetable}

\begin{deluxetable*}{llr@{$\pm$}lr@{$\pm$}lr@{$\pm$}lr@{$\pm$}l}
\tablecolumns{6}
\tablewidth{0pc}
\tablecaption{\obj\ Line Identifications \label{table:lines}}
\tablehead{
\colhead{Line} & \colhead{Date} & \multicolumn{2}{c}{Wavelength (\AA)} 
& \multicolumn{2}{c}{EW (\AA)} & \multicolumn{2}{c}{FWHM (\AA)} & \multicolumn{2}{c}{Velocity (${\rm km~s^{-1}}$)\tablenotemark{a}}}
\startdata
H$\alpha$ $\lambda6562.85$ & Feb 7.30 & 6565.0 & 1.6 & -74.7 & 4.8 & 54.1 & 3.4 & 2470 & 155\\[0pt]
 & Feb 12.21 & 6563.2 & 0.6 & -812.1 & 23.0 & 46.0 & 1.0 & 2100 & 45\\[0pt]
 & Feb 21.51 & 6564.7 & 0.0 & -1225.0 & 3.2 & 34.4 & 0.1 & 1570 & 5\\[0pt]
H$\beta$ $\lambda4860.36$ & Feb 7.30 & 4867.5 & 1.3 & -36.1 & 2.8 & 34.3 & 2.6 & 2115 & 160\\[0pt]
 & Feb 12.21 & 4861.6 & 0.7 & -296.4 & 14.0 & 31.2 & 1.4 & 1925 & 85\\[0pt]
 & Feb 21.51 & 4862.2 & 0.1 & -201.5 & 1.3 & 20.6 & 0.1 & 1275 & 5\\[0pt]
H$\gamma$ $\lambda4343.49$ & Feb 7.30 & 4347.7 & 1.6 & -19.8 & 2.0 & 30.1 & 3.6 & 2080 & 250\\[0pt]
 & Feb 12.21 & 4341.8 & 1.3 & -81.0 & 7.1 & 28.6 & 2.6 & 1975 & 175\\[0pt]
H$\delta$ $\lambda4101.77$ & Feb 12.21 & 4103.2 & 1.7 & -50.2 & 5.8 & 30.1 & 4.3 & 2200 & 315\\[0pt]
H$\epsilon$ $\lambda3970$ & Feb 12.21 & 3973.6 & 1.4 & -42.1 & 5.2 & 22.7 & 3.2 & 1715 & 240\\[0pt]
H$\zeta$ $\lambda3889$ & Feb 12.21 & 3891.1 & 1.8 & -19.1 & 3.9 & 17.8 & 6.4 & 1370 & 495\\[0pt]
Paschen$\iota$ $\lambda8750.47$ & Feb 7.30 & 8753.8 & 4.6 & -12.6 & 3.8 & 39.5 & 22.9 & 1350 & 785\\[0pt]
\ion{C}{1} $\lambda9111$ & Feb 7.30 & 9112.4 & 4.6 & -28.6 & 5.7 & 50.8 & 14.1 & 1675 & 465\\[0pt]
\ion{C}{1} $\lambda9408$ & Feb 7.30 & 9408.0 & 3.2 & -14.2 & 4.3 & 25.0 & 12.2 & 795 & 390\\[0pt]
\ion{C}{1} $\lambda9660$ & Feb 7.30 & 9669.3 & 11.4 & -12.0 & 6.2 & 38.7 & 30.4 & 1200 & 940\\[0pt]
\ion{N}{1}/\ion{Fe}{2} $\lambda7452$ & Feb 7.30 & 7414.4 & 8.1 & 8.9 & 3.0 & 55.5 & 36.1 & -1515 & 325\tablenotemark{b}\\[0pt]
& Feb 7.30 & 7476.1 & 7.1 & -7.5 & 3.0 & 40.6 & 30.9 & 970 & 285\tablenotemark{r}\\[0pt]
 & Feb 12.21 & 7465.8 & 7.9 & -25.4 & 8.3 & 48.2 & 24.7 & 1940 & 990\\[0pt]
 & Feb 21.51 & 7473.8 & 2.2 & -46.7 & 2.0 & 109.4 & 6.0 & 4400 & 240\\[0pt]
 \ion{N}{1} $\lambda8212$ & Feb 7.30 & 8168.7 & 6.0 & 19.1 & 4.1 & 59.4 & 18.0 & -1580 & 220\tablenotemark{b}\\[0pt]
 & Feb 7.30 & 8248.3 & 6.0 & -12.2 & 4.0 & 33.9 & 14.6 & 1325 & 220\tablenotemark{r}\\[0pt]
 & Feb 12.21 & 8231.0 & 6.8 & -53.1 & 10.0 & 78.8 & 24.8 & 2875 & 905\\[0pt]
 & Feb 21.51 & 8215.7 & 0.6 & -110.8 & 2.0 & 72.8 & 1.4 & 2660 & 50\\[0pt]
\ion{N}{1} $\lambda8692$/Ca~II & Feb 7.30 & 8648.7 & 9.6 & 13.4 & 4.4 & 51.8 & 26.1 & 1785 & 900\\[0pt]
\ion{N}{1} $\lambda8692$ & Feb 7.30 & 8720.2 & 4.3 & -10.3 & 3.2 & 34.1 & 23.6 & 1175 & 815\\[0pt]
 & Feb 12.21 & 8707.3 & 5.4 & -63.0 & 10.0 & 67.5 & 16.8 & 2330 & 580\\[0pt]
N~I 8692/Ca~II blend & Feb 21.51 & 8656.6 & 0.8 & -188.6 & 2.4 & \multicolumn{2}{l}{\nodata} & \multicolumn{2}{l}{\nodata}\\[0pt]
[\ion{O}{1}] $\lambda5577.34$ & Feb 21.51 & 5577.4 & 0.4 & -21.2 & 0.8 & 21.8 & 1.0 & 1170 & 50\\[0pt]
[\ion{O}{1}] $\lambda6300.30$ & Feb 21.51 & 6300.6 & 0.6 & -31.8 & 1.1 & 35.0 & 1.3 & 1665 & 60\\[0pt]
[\ion{O}{1}] $\lambda6363.78$ & Feb 21.51 & 6367.3 & 1.2 & -14.9 & 1.0 & 34.4 & 3.0 & 1620 & 140\\[0pt]
\ion{O}{1} $\lambda\lambda7773.75$ & Feb 7.30 & 7744.2 & 3.2 & 7.2 & 2.3 & 19.1 & 9.1 & -1140 & 125\tablenotemark{b}\\[0pt]
& Feb 7.30 & 7784.3 & 4.2 & -17.8 & 3.9 & 34.4 & 11.7 & 405& 160\tablenotemark{r}\\[0pt]
 & Feb 12.21 & 7780.8 & 1.9 & -220.3 & 16.0 & 53.0 & 3.5 & 2045 & 135\\[0pt]
 & Feb 21.51 & 7778.5 & 0.2 & -207.6 & 2.1 & 36.5 & 0.3 & 1410 & 15\\[0pt]
 \ion{O}{1} $\lambda\lambda8446.46$ & Feb 7.30 & 8410.0 & 9.8 & 5.6 & 3.2 & \multicolumn{2}{l}{\nodata} & \multicolumn{2}{l}{\nodata} \\[0pt]
 & Feb 7.30 & 8492.9 & 5.2 & -5.8 & 3.0 & 29.5 & 25.3 & \multicolumn{2}{l}{\nodata} \\[0pt]
 & Feb 12.21 & 8449.9 & 2.1 & -216.9 & 15.0 & 59.7 & 4.6 & 2120 & 160\\[0pt]
 & Feb 21.51 & 8447.5 & 0.1 & -668.6 & 3.2 & 43.5 & 0.2 & 1545 & 5\\[0pt]
\ion{O}{1} $\lambda9264$ & Feb 7.30 & 9268.1 & 3.6 & -22.8 & 5.3 & 30.5 & 9.1 & 985 & 295\\[0pt]
Na~I~D $\lambda\lambda5891.94$ & Feb 7.30 & 5868.7 & 4.6 & 8.0 & 2.6 & 21.0 & 13.1 & -1180 & 235 \tablenotemark{b}\\[0pt]
 & Feb 7.30 & 5906.5 & 6.3 & -4.5 & 2.2 & 23.4 & 17.4 & 740 & 320\tablenotemark{r}\\[0pt]
Na~I~D $\lambda6160$ & Feb 7.30 & 6131.1 & 11.5 & 2.4 & 2.2 & 24.1 & 20.8 & -1405 & 560\tablenotemark{b}\\[0pt]
& Feb 7.30 & 6167.6 & 4.1 & -7.1 & 2.4 & 21.9 & 11.6 & 370 & 200\tablenotemark{r}\\[0pt]
\ion{Ca}{2} $\lambda8498.02$ & Feb 7.30 & 8492.9 & 5.2 & -5.8 & 3.0 & 29.5 & 25.3 & 1040 & 895\\[0pt]
\ion{Ca}{2} $\lambda8542.09$ & Feb 7.30 & 8540.9 & 6.9 & -6.0 & 3.3 & \multicolumn{2}{l}{\nodata} & \multicolumn{2}{l}{\nodata}\\[0pt]
 & Feb 12.21 & 8557.0 & 3.9 & -28.6 & 7.3 & 35.0 & 17.8 & 1230 & 625\\[0pt]
\enddata
\tablecomments{Maximum light was on 2010 February \maxdate\ UT. All wavelengths reported are after de-redshifting all spectra using $z=0.00087$. Blue and red velocity-shifted P~Cygni components are measured from the rest-wavelength of the line species. Measured for where a nice fit was obtained, for low signal to line profiles, only equivalent width is reported. The central wavelengths of a line blend ($\lambda \lambda$) is the weighted average of the NIST reported line centers.} 
\tablenotetext{a}{Reported velocities use FWHM for pure emission lines only, and actual ejecta velocity is ${\rm FWHM}/2$. For P~Cygni line profiles, the velocity is demarcated as either the blue shifted or red shifted component from rest wavelength. A negative equivalent width corresponds to an emission line.}
\tablenotetext{b}{Blue Shifted P~Cygni Component.}
\tablenotetext{r}{Red Shifted P~Cygni Component.}
\end{deluxetable*}    

\end{document}